\documentclass[twocolumn]{aastex62}

\newcommand{\cone}{[\ion{C}{1}](1--0)~}

\usepackage{multirow}
\hypersetup{citecolor=blue,urlcolor=blue}
\received{June 18, 2018}
\revised{August 9, 2018}
\accepted{September 14, 2018}
\submitjournal{ApJ}

\shorttitle{Circumnuclear Multi-phase Gas in the Circinus Galaxy}
\shortauthors{T. Izumi et al.}

\begin{document}
\title{Circumnuclear Multi-phase Gas in the Circinus Galaxy II: \\
The Molecular and Atomic Obscuring Structures Revealed with ALMA}

\correspondingauthor{Takuma Izumi}
\email{takuma.izumi@nao.ac.jp}

\author[0000-0002-0786-7307]{Takuma Izumi}
\altaffiliation{NAOJ Fellow}
\affil{National Astronomical Observatory of Japan, 2-21-1 Osawa, Mitaka, Tokyo 181-8588, Japan}

\author{Keiichi Wada}
\affil{Kagoshima University, Kagoshima 890-0065, Japan}
\affil{Research Center for Space and Cosmic Evolution, Ehime University, Matsuyama 790-8577, Japan}
\affil{Hokkaido University, Faculty of Science, Sapporo 060-0810, Japan}

\author{Ryosuke Fukushige}
\affil{Kagoshima University, Kagoshima 890-0065, Japan} 

\author{Sota Hamamura}
\affil{Kagoshima University, Kagoshima 890-0065, Japan} 

\author{Kotaro Kohno}
\affil{Institute of Astronomy, Graduate School of Science, The University of Tokyo, 2-21-1 Osawa, Mitaka, Tokyo 181-0015, Japan}
\affil{Research Center for the Early Universe, Graduate School of Science, The University of Tokyo, 7-3-1 Hongo, Bunkyo, Tokyo 113-0033, Japan}

\begin{abstract}
We used the Atacama Large Millimeter/Submillimeter Array (ALMA) 
to map the CO(3--2) and [\ion{C}{1}](1--0) lines, as well as their underlying continuum emission, 
from the central $\sim 200$ pc region of the Circinus galaxy that hosts 
the nearest type 2 Seyfert-class active galactic nucleus (AGN), 
with a spatial resolution of $\sim 6-15$ pc. 
The lines and continuum-emitting regions consist of 
a circumnuclear disk (CND; 74 pc $\times$ 34 pc) and spiral arms. 
The distribution of the continuum emission revealed 
a temperature-dependent dust geometry and possibly polar dust elongation in the torus region. 
The molecular mass of the CND is $M_{\rm H2} \sim 3 \times 10^6~M_\odot$ 
with a beam-averaged H$_2$ column density of $\sim 5 \times 10^{23}$ cm$^{-2}$ toward the AGN position, 
which contributes significantly to the nuclear obscuration. 
The [\ion{C}{1}](1--0)/CO(3--2) ratio at the AGN position is unusually high, 
suggesting an X-ray dominated region-type chemistry. 
We decomposed the observed velocity fields into rotational and dispersion components, 
and revealed multi-phase dynamic nature in the $r \lesssim 10$ pc torus region, 
i.e., the diffuse atomic gas is more spatially extended 
along the vertical direction of the disk than the dense molecular gas. 
Through comparisons with our model predictions 
based on the radiation-driven fountain scheme, 
we indicate that atomic outflows are the driver of the geometrical thickness of the atomic disk. 
This supports the validity of the radiation-driven fountain scheme in the vicinity of this AGN, 
which would explain the long-lasting mystery, the physical origin of the AGN torus. 
\end{abstract}

\section{Introduction}\label{sec1}
The unified scheme of active galactic nuclei (AGNs) 
postulates that the observability of a broad line region (type 1 and 2) 
depends on the viewing angle of an optically and geometrically thick 
dusty/molecular torus that surrounds the central supermassive black hole \citep{1993ARA&A..31..473A,1995PASP..107..803U}. 
Spatially resolved thermal dust emission at near- to mid-infrared (NIR to MIR) wavelengths 
in AGNs indeed support the existence of compact ($< 10$ pc) obscuring structures 
\citep[e.g.,][]{2004Natur.429...47J,2013A&A...558A.149B,2014MNRAS.439.1648A}. 

Since the advent of the unification scheme, 
the physical origin of the geometrical thickness has been debated. 
In early theories, the torus was simplified as a continuous structure composed of dust 
and supported by, e.g., infrared (IR) radiation pressure and turbulence 
\citep[e.g.,][]{1992ApJ...401...99P,1993ApJ...418..673P,1994MNRAS.268..235G,2005A&A...437..861S}. 
The dust distribution was later revised to be clumpy in nature 
\citep[e.g.,][]{2002ApJ...570L...9N,2008ApJ...685..147N,2008ApJ...685..160N,
2006A&A...452..459H,2008A&A...482...67S,2012MNRAS.420.2756S}, 
which successfully reproduced the characteristic features 
of the AGN spectral energy distribution (SED) 
including the 9.7 $\mu$m silicate feature. 
Replacement of the torus by a magnetocentrifugally-driven 
disk wind containing dusty clumps was also proposed 
\citep[e.g.,][]{2006ApJ...648L.101E,2009ApJ...701L..91E}. 
Supported by the prevalence of circumnuclear starbursts around AGNs 
\citep[e.g.,][]{2004ApJ...617..214I,2007ApJ...671.1388D,2014ApJ...780...86E}, 
a supernova (SN)-driven turbulent torus model was also suggested 
\citep{2002ApJ...566L..21W,2009ApJ...702...63W}, 
where the SNe puffs up the disk gas and dust to form a toroidal structure 
at $\gtrsim 10$ parsec (pc) scales. 

However, these models are now challenged by the existence of 
{\it polar-elongation} in MIR continuum emission distributions 
revealed by high-resolution observations in nearby Seyfert galaxies 
\citep[e.g.,][]{2014A&A...563A..82T,2016ApJ...822..109A,2016A&A...591A..47L}, 
which seems to contradict the donut-like dust geometry in the equatorial plane postulated in the unified scheme. 
Recent detailed radiative transfer simulations have revealed that a dusty hollow cone 
illuminated by the central engine can explain the observed morphology \citep{2017MNRAS.472.3854S}. 
Therefore, it is noteworthy that the radiation-driven fountain model \citep{2012ApJ...758...66W}, 
in which the circulation of AGN-driven dusty outflows, failed-winds, and inflows 
jointly form a geometrically thick structure, 
can naturally reproduce such a dusty hollow cone 
and the polar elongation at MIR wavelengths \citep{2014MNRAS.445.3878S}. 

\begin{figure}
\begin{center}
\includegraphics[scale=0.42]{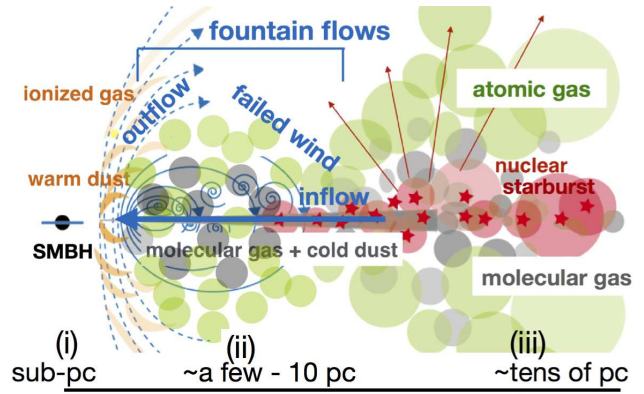}
\caption{
A schematic picture of our multi-phase dynamic torus model. 
There are three key regions/structures, 
(i) low-density hot and dusty outflows at a sub-pc region, 
(ii) low-density cold dusty outflows and a turbulent 
(i.e., geometrically thick) region induced 
by the shocks due to failed winds from a region of a few pc to 10 pc, 
and (iii) a geometrically thin disk at a $\gtrsim$10 pc region, 
where gas and dust can be puffed-up if supernova feedback is sufficient. 
}
\label{fig1}
\end{center}
\end{figure}

This fountain model, which was later complemented by the aforementioned SN feedback process 
and non-equilibrium X-ray dominated region \citep[XDR, e.g.,][]{1996ApJ...466..561M} chemistry \citep{2016ApJ...828L..19W}, 
predicts that the gas and dust of a $\lesssim$ tens of pc scale circumnuclear disk 
\citep[CND;][]{2016ApJ...827...81I} are highly dynamic and non-static, 
and consist of three main regions/structures that replace the classic torus: 
(i) low density hot (= \ion{H}{2}) and dusty outflows that are launched 
due to X-ray heating and radiation pressure at a sub-pc region 
(the outflows irradiated by the central source form narrow line regions; K. Wada et al. in preparation), 
(ii) low density cold (= \ion{H}{1}) dusty outflows and failed winds\footnote{A part of the outflow that falls back to the disk due to disk and SMBH gravities.} 
at a region of a few pc to 10 pc that cause a mid-plane disk (= \ion{H}{1} + H$_2$) 
to become highly turbulent and then geometrically thick, 
and (iii) a geometrically thin disk (= \ion{H}{1} + H$_2$) located at a region of $\gtrsim 10$ pc. 
SN-driven turbulence, if significant, can cause this latter region to become geometrically thick. 
This model is hereafter denoted as the {\it multi-phase dynamic torus model} (Figure \ref{fig1}). 
The inflow gas passes through the dense mid-plane of the disk. 
Thus, the circulation of the inflows, outflows, and failed winds jointly form the {\it fountain}. 
Components (ii) and (iii) are basically responsible for the nuclear obscuration, 
with significant substructures in the column density \citep{2015ApJ...812...82W}. 
In either region (ii) or (iii), diffuse atomic gas is spatially 
more extended along the vertical direction of the disk than dense molecular gas, 
due to complex interplay between the gas dynamics, AGN radiation, and local heating/cooling. 
For example, in region (ii), spatially extended outflows 
are preferentially observed in the atomic gas rather than molecular gas in this model. 
Thus, high-resolution observations of the multi-phase interstellar medium (ISM) 
around AGNs are essential to test this picture of the multi-phase gas circulation. 

The Atacama Large Millimeter/submillimeter Array (ALMA) has the capability to perform such observations. 
Indeed, it has finally begun to detect cold molecular gas and dust emission 
from the central $\lesssim 10$ pc region of nearby AGNs 
including NGC 1068 \citep{2016ApJ...822L..10I,2018ApJ...853L..25I,2016ApJ...823L..12G,2016ApJ...829L...7G}, 
NGC 1377 \citep{2017A&A...608A..22A}, 
Centaurus A \citep{2017ApJ...843..136E}, 
NGC 1097 \citep{2017ApJ...845L...5I}, 
and NGC 5643 \citep{2018arXiv180404842A}. 
However, as these studies observed dense molecular gas only, 
the physical and chemical nature of multi-phase gas remain unclear. 
This situation could be improved by using the atomic carbon line 
[\ion{C}{1}]($^{3}P_1$--$^{3}P_0$) $\equiv$ \cone that is now observable with ALMA Band 8 
for the case of nearby galaxies \citep{2016A&A...592L...3K,2018arXiv180200486M}. 
Contrary to the prediction of classic chemical models that \ion{C}{1} is distributed in a thin layer 
between the ionized part and the shielded molecular part of a cloud \citep[e.g.,][]{1999RvMP...71..173H}, 
it is now well known that \ion{C}{1} is actually concurrent through the cloud 
\citep[e.g.,][]{2001ApJ...548..253O,2002ApJS..139..467I,2013ApJ...774L..20S}. 
This is likely due to the different formation timescale between 
\ion{C}{1} and CO \citep{1992ApJ...392..551S}, 
and/or strong interstellar turbulence \citep{2015MNRAS.448.1607G}. 
It is noteworthy that in active environments with strong radiation fields, 
such as an XDR \citep[e.g.,][]{1996ApJ...466..561M}, 
C-bearing species such as CO are dissociated and enhance the abundance of \ion{C}{1}. 
Thus, \ion{C}{1} lines will be excellent probes of AGN signatures. 

In this work, we present our high-resolution ($< 1\arcsec$) ALMA Cycle 4 observations 
of the CO(3--2) ($\nu_{\rm rest}$ = 345.7960 GHz) and 
the \cone ($\nu_{\rm rest}$ = 492.1607 GHz), 
as well as the underlying continuum emission toward 
the central $\sim 200$ pc region of the Circinus galaxy 
(hereafter, denoted as Circinus). 
Our aim is to investigate the nature of the circumnuclear obscuring material, 
by focusing particularly on the differences in the dynamical 
or geometrical structures as a function of the phase of the gas (molecular vs. atomic). 
The observed properties will be compared to our multi-phase dynamic torus model, 
where key parameters (e.g., black hole mass, Eddington ratio, gas mass) 
were matched to those measured in Circinus, as well as post-processed 
non-local thermodynamic equilibrium (non-LTE) 
radiative transfer calculations were performed 
for multiple CO lines \citep[our Paper-1,][]{2018ApJ...852...88W} 
and \ion{C}{1} lines (this work). 

\subsection{Our target: The Circinus galaxy}
Circinus is a spiral galaxy \citep[SAb;][]{1991rc3..book.....D} 
with a high inclination angle on the galactic scale \citep[$\sim 65\arcdeg$;][]{1977A&A....55..445F}, 
and hosts the nearest \citep[$D$ = 4.2 Mpc, 1$\arcsec$ = 20.4 pc;][]{2009AJ....138..323T} 
type 2 Seyfert nucleus \citep{1994A&A...288..457O,1996A&A...315L.109M}. 
It shows a one-sided ionization cone and outflowing gas that extends up to kpc scales 
in H$\alpha$, [\ion{O}{3}], and several other coronal emission lines 
\citep[e.g.,][]{1994Msngr..78...20M,1997ApJ...479L.105V,2006A&A...454..481M}. 
The genuine existence of an obscured type 1 nucleus was confirmed 
by the spectropolarimetric detection of a broad H$\alpha$ line \citep[$\simeq 3300$ km s$^{-1}$;][]{1998A&A...329L..21O}. 
Circinus is also undergoing a modest star-forming activity, 
as measured using far-infrared luminosity \citep[star formation rate (SFR) $\sim$a few $M_\odot$ yr$^{-1}$;][]{1984A&A...135..281M,1998MNRAS.300.1119E} 
and several hydrogen recombination lines \citep[e.g.,][]{1994Msngr..78...20M,2000AJ....120.1325W}, 
while the $\lesssim 100$ pc scale SFR is admittedly low \citep[$\sim 0.1~M_\odot$ yr$^{-1}$;][]{2014ApJ...780...86E}: 
thus, significant SN feedback might not be expected there. 

As is typical in the case of a late-type galaxy, a huge amount of 
molecular gas ($\gtrsim 10^{8-9}~M_\odot$) was detected in Circinus through observations of multi-$J$ CO lines 
including isotopologues, as well as other dense gas tracers such as HCN and HCO$^+$ 
\citep[e.g.,][]{1991A&A...249..323A,1992A&A...265..487I,1998MNRAS.300.1119E,1998A&A...338..863C,
1999A&A...344..767C,2001A&A...367..457C,2008MNRAS.389...63C,2008A&A...479...75H,2014A&A...568A.122Z,2015A&A...578A..95I}. 
While the intense \ion{H}{1} emission is distributed in a $\sim 10$ kpc-radius region 
dominated by large-scale spiral arms \citep[][]{1999MNRAS.302..649J}, 
molecular gas is concentrated in the inner $\sim 1$ kpc-radius region 
\citep{1998A&A...338..863C,2008MNRAS.389...63C,1998MNRAS.300.1119E}. 
An expected morphology for that molecular gas structure, 
with the H$_2$ mass ($M_{\rm H_2}$)
\footnote{The standard Galactic CO-to-H$_2$ 
conversion factor of $\sim 2 \times 10^{20}$ cm$^{-2}$ (K km s$^{-1}$)$^{-1}$ 
is assumed for most of the previous $M_{\rm H_2}$ measurement, 
although \citet{2014A&A...568A.122Z} suggested a factor of $\sim 5$ lower value 
in Circinus through the physical modeling of multi-$J$ CO emission lines.} 
of $\sim 4 \times 10^8~M_\odot$ \citep[$r < 440$ pc;][]{1998MNRAS.300.1119E}, 
has been a ring/disk-like one plus widespread outflows perpendicular 
to the morphological major axis of this galaxy \citep{1998A&A...338..863C,1999A&A...344..767C}. 
However, recent $\sim 2\arcsec-3\arcsec$ resolution CO(1--0) mapping with ALMA 
imaged instead a more spiral arm-like gas distribution \citep{2016ApJ...832..142Z}, 
while revealing another molecular outflow at 35$\arcsec$ northwest of the nucleus. 

With regard to the nuclear scale, X-ray spectra below 10 keV 
exhibit flat continuum and prominent 6.4 keV Fe K$\alpha$ line emission, 
indicative of strong Compton scattering \citep{1996MNRAS.281L..69M}. 
Subsequent harder X-ray observations at $>30$ keV confirmed the existence of 
the Compton-thick nucleus \citep[e.g.,][]{1999A&A...341L..39M,1999MNRAS.310...10G}, 
with a line-of-sight obscuring column density of $N_{\rm H} = (6-10) \times 10^{24}$ cm$^{-2}$, 
and an absorption-corrected 2--10 keV intrinsic luminosity of $L_{\rm 2-10keV} = (2.3-5.1) \times 10^{42}$ erg s$^{-1}$, 
respectively \citep{2014ApJ...791...81A}. 
Detections of H$_2$O mega-masers at both mm and submm wavelengths at the heart of Circinus 
support the idea that a dense Keplerian disk is located there \citep{1997ApJ...474L.103G,2003ApJ...590..162G,2013ApJ...768L..38H}, 
with the mass of the central supermassive black hole (SMBH) given as 
$M_{\rm BH} = (1.7\pm0.3) \times 10^6~M_\odot$ \citep{2003ApJ...590..162G}. 
With this $M_{\rm BH}$, \citet{2007A&A...474..837T} 
estimated the Eddington ratio of the AGN as $\sim 0.2$. 

Our prime reason for studying Circinus is that it clearly exhibits pc-scale polar elongation 
at MIR continuum emission \citep{2014A&A...563A..82T}, 
which makes it an ideal laboratory to test 
our multi-phase dynamic torus model. 
We describe the ALMA Cycle 4 observations in \S~2. 
The continuum emission maps are shown in \S~3, 
while line emission distributions and their ratio are reported in \S~4. 
Details of the gas kinematics of both the CO(3--2) and the \cone lines are presented in \S~5. 
We compare the observed torus properties 
with the predictions of our multi-phase dynamic torus model in \S~6. 
Finally, our conclusions are summarized in \S~7.

\section{Observations and Data Analysis}\label{sec2} 
\subsection{ALMA observations}
We observed Circinus with ALMA Band 7 and 8 during 2016--2017 
using 42--47 antennas, as a Cycle 4 program (ID = 2016.1.01613.S, PI = T. Izumi). 
Table \ref{tbl1} summarizes the log of our observations. 
Observations were conducted in a single pointing with fields of view 
of 18$\arcsec$ (Band 7) and 13$\arcsec$ (Band 8), 
which fully covered the central $\sim 2\arcsec$ of the CND (see \S~4). 
The expected maximum recoverable scales per pointing are 
$\sim 7\arcsec$ (Band 7) and $\sim 5\arcsec$ (Band 8). 
We set the phase tracking center to ($\alpha_{\rm J2000.0}$, $\delta_{\rm J2000.0}$) 
= (14$^{\rm h}$13$^{\rm m}$09$^{\rm s}$.950, $-$65$\arcdeg$20$\arcmin$21$\arcsec$.0), 
which is one of the nuclear 22 GHz H$_2$O maser spots of Circinus \citep{2003ApJ...590..162G}. 
The angular separations between this tracking center and the phase calibrators are $\sim 3\arcdeg$. 
In both Band 7 and 8 observations, one of the four spectral windows 
(each with a width of 1.875 GHz) was used to 
fully cover the CO(3--2) or [\ion{C}{1}](1--0) emission lines, 
both in the 2SB dual-polarization mode. 

Data reduction, calibration, and analyses were performed with CASA version 4.7 
\citep{2007ASPC..376..127M} in the usual way. 
The line and underlying continuum emission were 
reconstructed using the \verb|CLEAN| task with Briggs weighting (robust = 0.5). 
The velocity spacings of the original data were 3.4 km s$^{-1}$ (Band 7) 
and 2.4 km s$^{-1}$ (Band 8) per channel, 
but several channels were binned to improve the signal-to-noise ratio, 
which resulted in a final and common velocity resolution of $\sim 10$ km s$^{-1}$. 
Note that velocities are expressed in the optical convention in this work 
(local standard of rest [LSR] frame). 

The achieved synthesized beams and rms sensitivities are listed in Table \ref{tbl2}. 
The rms values in the line cubes were measured 
at channels free of line emission (i.e., thermal noise), 
whereas those of the continuum maps were measured in areas free of such emission. 
These continuum emission were subtracted in the $uv$ plane before making the line cubes. 
Throughout this paper, the pixel scale of Band 7 maps is set to 0$\arcsec$.03, 
whereas that of Band 8 maps is 0$\arcsec$.1, 
and emission with $<1.5\sigma$ are clipped to enhance the clarity of the images. 
The kinematic position angle (PA) of the major axis is defined to be on the receding half of the galaxy, 
taken anti-clockwise from the north direction on the sky. 
Given this definition, we added 180$\arcdeg$ 
to some morphological PAs reported in previous works to maintain consistency. 
The absolute flux uncertainty is $\sim 10\%$ 
according to the ALMA Cycle 4 Proposer's Guide, 
but the displayed errors indicate only statistical ones unless otherwise mentioned. 
Some parts of our analyses were also performed with the MIRIAD package \citep{1995ASPC...77..433S}.

\begin{table*}
\begin{center}
\caption{Log of our ALMA observations \label{tbl1}}
\begin{tabular}{ccccccccc}
\tableline\tableline
\multirow{2}{*}{Line} & Date & \multirow{2}{*}{Number of Antennae} & Baseline & On-source time & \multicolumn{3}{c}{Calibrator} & $T_{\rm sys}$ \\ \cline{6-8}
 & (UT) &  & (m) & (min) & Band pass & Gain & Flux & (K) \\ 
 \tableline
 \multirow{3}{*}{CO(3--2)} & 2016 Nov 24 & 42 & 15--704 & 42.5 & J1427-4206 & J1424-6807 & J1427-4206 & $\sim 130$ \\
  & 2016 Nov 26 & 42 & 17--704 & 42.5 & J1427-4206 & J1424-6807 & J1617-5848 & $\sim 140$ \\ 
  & 2017 May 5 & 47 & 15--1124 & 42.5 & J1427-4206 & J1424-6807 & J1427-4206 & $\sim 140$ \\ \hline
 $[$\ion{C}{1}$]$(1--0) & 2017 Mar 18 & 43 & 15--287 & 31.5 & J1427-4206 & J1424-6807 & J1427-4206 & $\sim 500$ \\ 
\tableline
\end{tabular}
\end{center}
\end{table*}

\begin{table*}
\begin{center}
\caption{Achieved cube parameters \label{tbl2}}
\begin{tabular}{cccccc}
\tableline\tableline
\multirow{2}{*}{Emission} & $\nu_{\rm rest}$ & Beam & Beam & rms & Peak \\ 
 & (GHz) & ($\arcsec \times \arcsec$) ($\arcdeg$) & (pc $\times$ pc) & (mJy beam$^{-1}$) & (mJy beam$^{-1}$) \\
\tableline
CO(3--2) & 345.796 & 0.29 $\times$ 0.24 (153.6) & 5.9 $\times$ 4.9 & 0.37 & 301 \\
Band 7 continuum & 351 & 0.29 $\times$ 0.24 (156.0) & 5.9 $\times$ 4.9 & 0.08 & 22.4 \\ 
$[$\ion{C}{1}$]$(1--0) & 492.161 & 0.71 $\times$ 0.66 (95.9) & 14.5 $\times$ 13.4 & 3.10 & 1435 \\
Band 8 continuum & 485 & 0.71 $\times$ 0.65 (94.2) & 14.5 $\times$ 13.2 & 0.80 & 87.8 \\ \hline
\tableline
\end{tabular}
\tablecomments{In the line cubes, rms sensitivities indicate the values at a velocity resolution of 10 km s$^{-1}$.}
\end{center}
\end{table*}

\subsection{Ancillary data}
We retrieved the {\it Hubble Space Telescope (HST)} Planetary Camera 
images (PC-F656N, F606W, F814W) of Circinus from the Hubble Legacy Archive\footnote{\url{https://hla.stsci.edu}}. 
The images were calibrated using the {\it HST} pipeline. 
Continuum emission was subtracted from the F656N image using 
the adjacent continuum following standard line extraction procedures 
to construct an image of the H$\alpha$ emission line. 
The structures seen in the resultant image are consistent 
with those found in previous works \citep[e.g.,][]{1994Msngr..78...20M,2000AJ....120.1325W}. 
We also use the $Ks$-band image obtained with the NaCo/VLT \citep{2016MNRAS.457L..94M} 
to compare the spatial distributions of the continuum emission.

\section{Continuum Emission}\label{sec3}
\subsection{Spatial distributions}\label{sec3.1}
\begin{figure*}
\begin{center}
\includegraphics[scale=0.53]{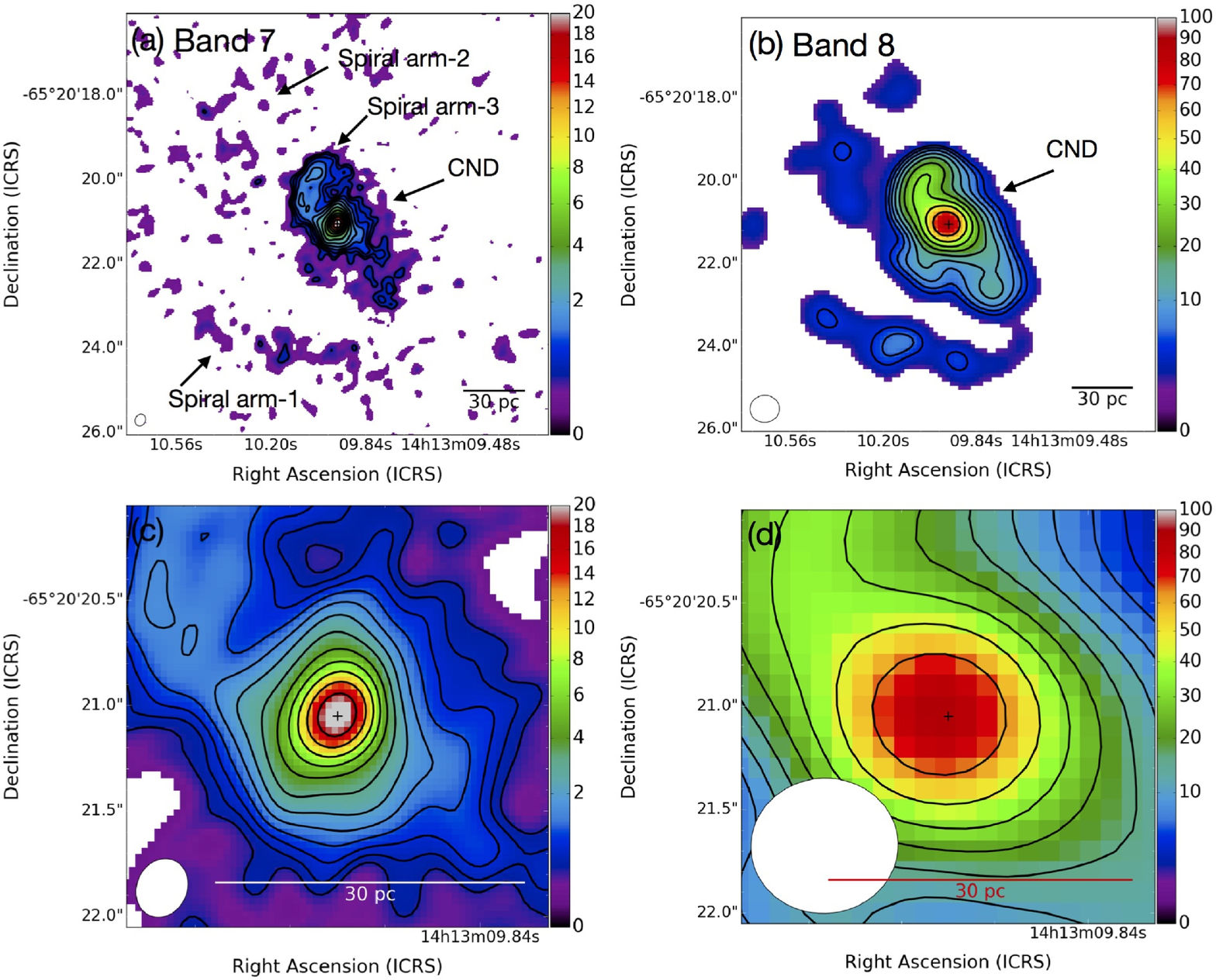}
\caption{
(a) Band 7 ($\nu_{\rm rest}$ = 351 GHz, $\lambda_{\rm rest}$ = 854 $\mu$m) continuum emission 
in the central $10\arcsec \times 10\arcsec$ region of the Circinus galaxy, 
shown in both color scale (mJy beam$^{-1}$) 
and contours ($5 \times 2^{(n-1)/2}~\sigma$ with $n$ = 1, 2, $\dots$, and 12, 
where 1$\sigma$ = 0.075 mJy beam$^{-1}$). 
(b) Band 8 ($\nu_{\rm rest}$ = 485 GHz, $\lambda_{\rm rest}$ = 618 $\mu$m) continuum emission 
in the same region as shown in (a). 
Contours indicate $5 \times 2^{(n-1)/2}~\sigma$ with $n$ = 1, 2, $\dots$, and 9, 
where 1$\sigma$ = 0.80 mJy beam$^{-1}$. 
(c) Zoomed-in view of the central $2\arcsec \times 2\arcsec$ of (a). 
(d) Zoomed-in view of the central $2\arcsec \times 2\arcsec$ of (b). 
In each panel, the central plus sign marks the AGN position 
and the synthesized beams are shown as a white ellipse. 
These maps were constructed with Briggs weighting (robust = 0.5), 
without correcting the primary beam attenuations. 
}
\label{fig2}
\end{center}
\end{figure*}

Figure \ref{fig2} shows the spatial distributions 
of the ALMA Band 7 ($\nu_{\rm rest}$ = 351 GHz or $\lambda_{\rm rest}$ = 854 $\mu$m) 
and 8 ($\nu_{\rm rest}$ = 485 GHz or $\lambda_{\rm rest}$ = 618 $\mu$m) continuum emission 
at the central $10\arcsec \times 10\arcsec$ ($\sim$ 200 pc $\times$ 200 pc) 
and $2\arcsec \times 2\arcsec$ ($\sim$ 40 pc $\times$ 40 pc) regions of Circinus. 
The total fluxes are $\sim 66.5$ mJy (Band 7) and $\sim 320$ mJy (Band 8) within the 10$\arcsec$ box. 
Each of the emission peaks at 
$\alpha_{\rm ICRS}$ = 14$^{\rm h}$13$^{\rm m}$09$^{\rm s}$.948, and 
$\delta_{\rm ICRS}$ = $-$65$\arcdeg$20$\arcmin$21$\arcsec$.05, 
which coincides with the AGN position identified 
by the very long baseline interferometry (VLBI) 
H$_2$O maser observations \citep[$\alpha_{\rm J2000.0}$ = 14$^{\rm h}$13$^{\rm m}$09$^{\rm s}$.953, 
$\delta_{\rm J2000.0}$ = $-$65$\arcdeg$20$\arcmin$21$\arcsec$.187;][]{2003ApJ...590..162G} within positional uncertainties. 
Hereafter, we define our continuum peak position as the AGN location. 
The peak flux densities are 22.4 mJy beam$^{-1}$ (Band 7) 
and 87.8 mJy beam$^{-1}$ (Band 8), respectively. 

Although the achieved angular resolutions were admittedly different (Table \ref{tbl2}), 
consistent spatial distributions can be identified between the Band 7 and 8 maps. 
The continuum-emitting regions appear to consist of (likely) three spiral arms 
which converge to the central bright CND or jointly constitute the CND: 
these structures are better echoed by the CO(3--2) and $[$\ion{C}{1}$]$(1--0) 
spatial distribution maps (see also Figure \ref{fig6} as a reference). 
Note that we were unable to find a clear indication of 
the nuclear gaseous bar postulated by \citet{2000ApJ...531..219M} southeast, 
nor an expected counter bar northwest, of the nucleus. 
This may imply that an aligned configuration of dense molecular clouds in the spiral arms (\S~4), 
viewed with a high inclination might have been mistaken as a bar. 

\subsection{Physical nature of the continuum emission}\label{sec3.2}
The nature of the observed submm continuum emission is of interest 
because the obscuring structure of this AGN would begin to be directly traced at the high resolutions obtained here. 
Although this galaxy possesses a radio jet \citep{1990MNRAS.244..130H,1998MNRAS.297.1202E,2010MNRAS.402.2403M}, 
we first argue that contamination by such synchrotron emission 
is not significant at the ALMA Band 7 and 8 frequencies. 
For example, a Band 7 flux density at the central 1$\arcsec$.4 region expected by 
extrapolating the $\lambda = 6$ cm emission (50 mJy) with a typical synchrotron spectral 
index\footnote{Flux (synchrotron) $\propto \nu^\alpha$.} of $\alpha = -0.7$ 
is only $\sim 2.5$ mJy, which is $\sim 9$ times smaller than 
the value measured at a much smaller beam 
($0\arcsec.29 \times 0\arcsec.24$) placed at the AGN position. 
Note that \citet{1998MNRAS.297.1202E} revealed a very flat radio spectral index 
at the nuclear region of Circinus at $\nu_{\rm rest} \lesssim 8$ GHz 
primarily due to optically thick nature at that frequency range. 
However, it is more reasonable to adopt the above $\alpha = -0.7$ for higher frequencies, 
where synchrotron emission becomes optically thinner. 

A Band 8 to 7 flux density ratio at the AGN position, 
after matching the Band 7 $uv$ range and beam size to the Band 8 ones 
($uv$ range $\simeq$ 25--470 k$\lambda$; the resultant Band 7 flux density = 26.3 mJy beam$^{-1}$), 
revealed a steep spectral index of 3.72, or dust emissivity index
\footnote{Flux (thermal dust) $\propto \nu^{2+\beta}$.} of $\beta = 1.72$. 
This is consistent with a typical $\beta$ index observed in 
local star-forming galaxies \citep[$\sim 1.8$;][]{2013MNRAS.433..695C}. 
Thus, the nuclear submm SED of Circinus is dominated by thermal dust emission. 
We found that the Band 8 continuum flux density at the AGN position is almost consistent with 
the value expected by the $T_{\rm dust} \sim 300$ K black body model in \citet[][]{2014A&A...563A..82T}, 
which was introduced to describe the observed IR SED at the (circum)nuclear region. 
Meanwhile, the Band 7 flux density at the same position 
falls by a factor of a few below the model prediction, 
which seems to reflect the fact that the dust continuum emission 
tends to follow the modified black body spectrum more, 
rather than the black body model, at a longer wavelength.

\subsection{Relative locations}\label{sec3.3}
\begin{figure}[h]
\begin{center}
\includegraphics[scale=0.45]{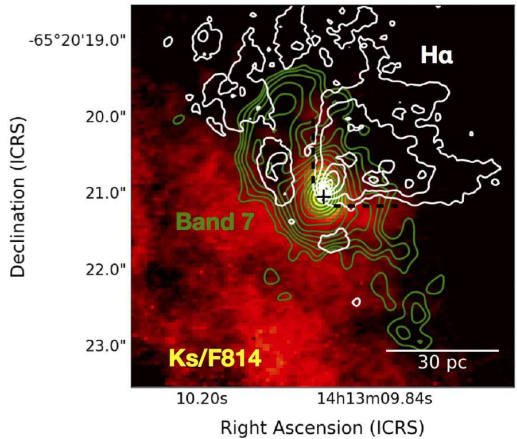}
\caption{
The overlay of the Band 7 continuum map (as shown in Figure \ref{fig2}; green contours) 
on the $Ks$/F814 flux ratio map \citep[color scale; brighter regions denote those with higher ratios,][]{2016MNRAS.457L..94M}. 
The $Ks$/F814 ratio distribution around the AGN position delineates the boundary of the $V$-shaped 
H$\alpha$ emission line distribution (white contours; 0.1, 0.2, 0.4, 0.6, $\cdots$, and 1.0 in units of counts s$^{-1}$ pixel$^{-1}$). 
The spiral-like structure seen in the Band 7 continuum emission 
traces the radially farther region than the very nuclear 
warm dust structure seen in the $Ks$/F814 ratio map (black dashed lines). 
}
\label{fig3}
\end{center}
\end{figure}

It is remarkable that the extended spiral arm-like or filamentary structure 
seen around the CND in our submm continuum maps 
traces the radially farther region from the center 
than the nuclear bright V-shaped structure seen in the $Ks$/F814 flux ratio map 
(Figure \ref{fig3}: the latter seems to collimate the ionized cone 
seen in the H$\alpha$ line emission distribution) 
that enhances the distribution of warm dust emission \citep{2016MNRAS.457L..94M}. 
This indicates the existence of temperature-dependent dusty structures around this AGN. 
This relative geometry, viewed with a high inclination angle 
\citep[$\sim 65\arcdeg$--$75\arcdeg$,][]{1977A&A....55..445F,1998MNRAS.300.1119E,2014A&A...563A..82T}, 
is also consistent with the prediction of our multi-phase dynamic torus model 
where the warm dust is embedded in outflows at the inner part of the disk, 
and the cold dust is located in the extended disk \citep[][see also Figure \ref{fig1}]{2014MNRAS.445.3878S,2016ApJ...828L..19W}. 
Within the framework of our model, the nuclear warm dust 
seen in the $Ks$/F814 ratio map is thus considered to be distributed 
in a geometrically thick volume, whereas the cold dust is in a thinner structure. 
This warm dust carries only $\sim 1/3$ of the obscuring matter required 
to make a type 1 Seyfert nucleus appear as an obscured 
type 2 nucleus such as Circinus \citep{2016MNRAS.457L..94M}. 
Thus, an additional absorber is required to completely obscure the nucleus: 
we suggest that the extended cold material in the CND studied here contributes to that obscuration (see also \S~4)
\footnote{We do not estimate dust mass here based on our continuum measurements 
because the temperature $T_{\rm dust}$ of the extended cold dust component, 
which would be a spatially different structure 
from the warm dust component ($\sim 300$ K black body), is unknown at present.}.

\subsection{The polar elongation}\label{sec3.3}
\begin{figure}[h]
\begin{center}
\includegraphics[scale=0.4]{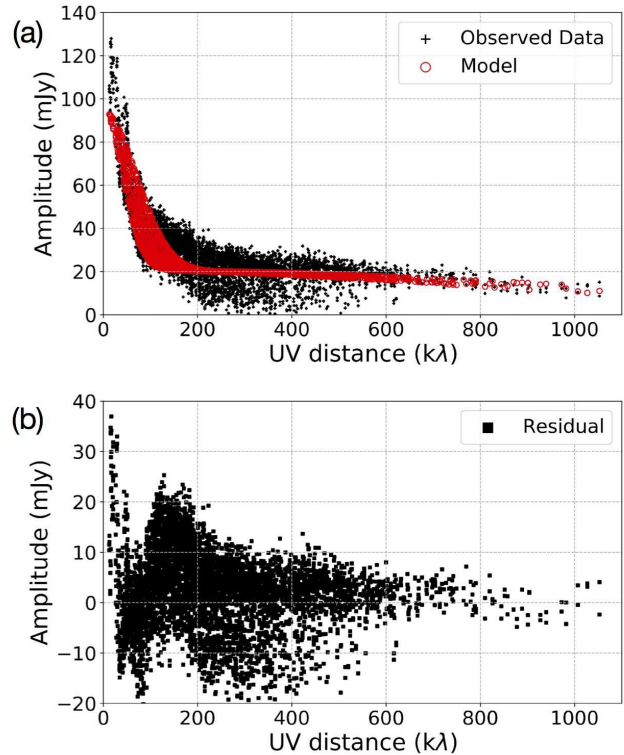}
\caption{
(a) ALMA $uv$ distance vs amplitude plot of the Band 7 continuum data of the Circinus galaxy (black). 
The model visibility data of our double Gaussian fit, returned by the UVMULTIFIT task, is also shown (red). 
(b) Residual visibility amplitude of the Band 7 continuum data, 
after subtracting the model component, shown as a function of $uv$ distance. 
}
\label{fig4}
\end{center}
\end{figure}

To further investigate the detailed spatial distribution of the submm continuum emission, 
we applied the \verb|UVMULTIFIT| procedure \citep{2014A&A...563A.136M} 
to the Band 7 continuum visibility data, 
in which we have sampled long $uv$ distance visibilities denser than the case of the Band 8 data. 
This model fitting method to the direct interferometric observable (visibility) 
is preferable to avoid systematic uncertainty in the image-plane fitting due to, 
e.g., non-linear deconvolution algorithm, particularly when we fit beam-unresolved components. 

The achieved visibility data is shown in Figure \ref{fig4}a as a function of $uv$ distance (UVD), 
after averaging the visibilities over the observation time and frequency (representative frequency = 350.2 GHz). 
The almost constant amplitude at the UVD $\gtrsim 600$ k$\lambda$ manifests 
the existence of a compact component, 
while the short UVD components reflect spatially extended structures. 
Here we performed a two-dimensional {\it double} Gaussian fit to this visibility data: 
the use of double components is motivated by 
the existence of a pc-scale MIR polar elongation 
\citep{2007A&A...474..837T,2014A&A...563A..82T}, 
in addition to the extended CND seen in Figure \ref{fig2}. 
The results of the fit are summarized in Table \ref{tbl3} 
and the corresponding model is displayed in Figure \ref{fig4}a as well. 
As expected in the above, our best fit indeed supports the existence 
of a compact component at the AGN position (Component-1 in Table \ref{tbl3}) 
in addition to the apparently extended structure (Component-2). 
The latter would correspond to a bright part of the CND (see also Figure \ref{fig2}). 
The fact that Component-2 is brighter than Component-1 
is consistent with the prediction of our multi-phase dynamic torus model 
that cold dust visible at longer FIR to submm wavelengths 
is predominantly located in an extended disk, 
while more centrally concentrated components including warm dusty outflows 
are prominent at shorter NIR to MIR wavelengths \citep{2014MNRAS.445.3878S}. 
Note that, however, we also found that the residual amplitude remains high 
at the short UVD ($\lesssim 400$ k$\lambda$; Figure \ref{fig4}b). 
This is due to the fact that there are some other 
spatially extended circumnuclear dusty structures in Circinus (e.g., spiral arms; Figure \ref{fig2}). 
As we would like to focus on the very nuclear structures in this subsection, 
fitting all components is beyond our scope. 

\begin{figure}[h]
\begin{center}
\includegraphics[scale=0.25]{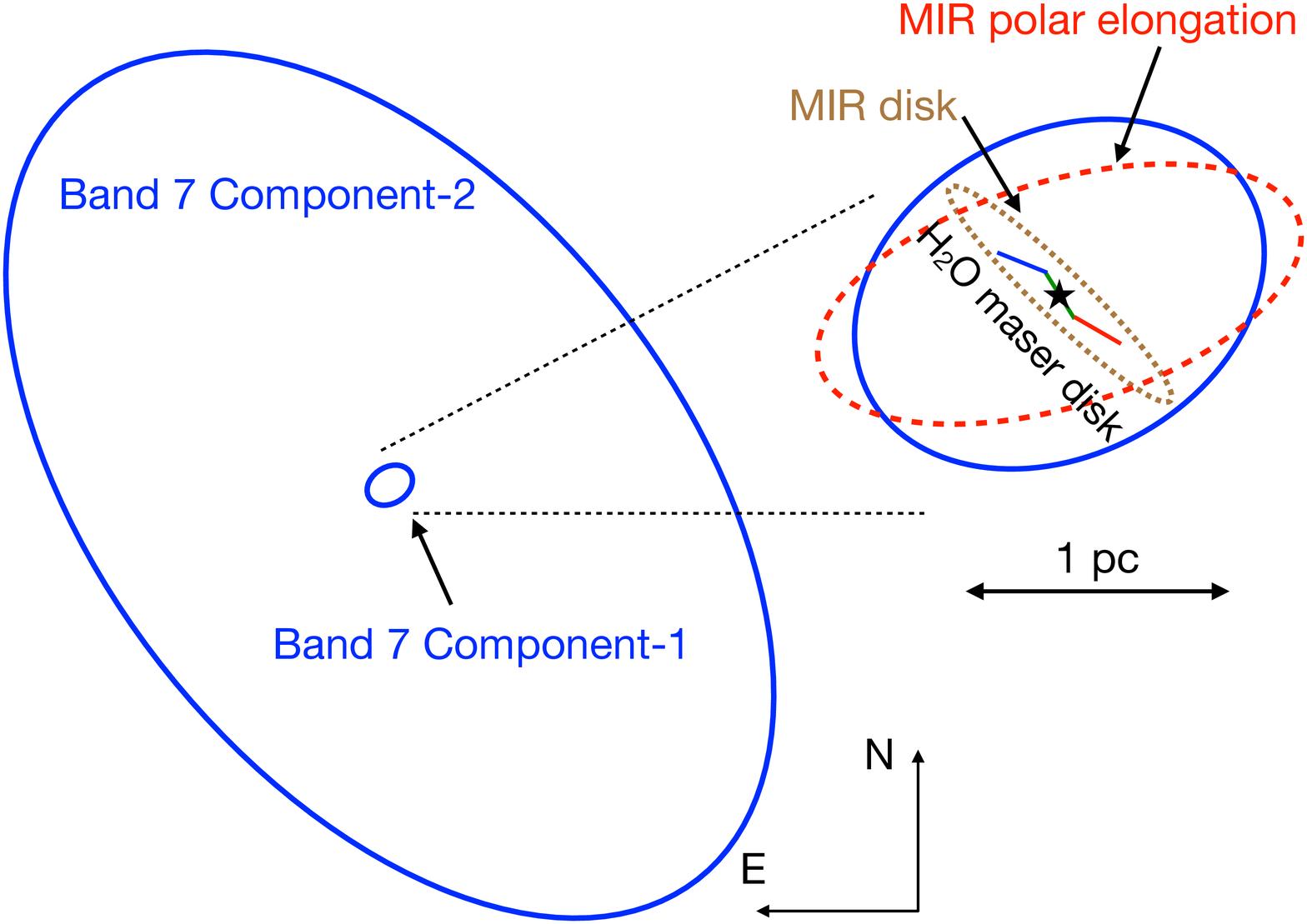}
\caption{
Schematic picture of the nuclear obscuring structures in the Circinus galaxy. 
Components 1 and 2 revealed by our ALMA Band 7 continuum observations 
are denoted by the blue solid line ellipses. 
The central $\sim 1$ pc region is zoomed-in on the right side. 
Also plotted are the MIR disk (brown short-dashed line ellipse), 
the MIR polar elongation \citep[red dashed line ellipse,][]{2014A&A...563A..82T}, 
and the 22 GHz H$_2$O maser disk \citep[central solid line, colors represent 
the line-of-sight velocity structure,][]{2003ApJ...590..162G}, respectively. 
}
\label{fig5}
\end{center}
\end{figure}

The spatial distributions of several (circum-)nuclear 
major components that we are interested in are illustrated in Figure \ref{fig5}. 
The axis ratio of this Component-2 suggests that this structure 
is moderately inclined if it is a circular disk ($\simeq 55\arcdeg$). 
This inclination angle is slightly smaller than the galactic-scale one ($\simeq 65\arcdeg$) 
possibly because the spatially extended/elongated emission along 
the northeast-southwest direction, particularly at the southwestern side 
of the nucleus, is faint (see also Figure \ref{fig2}a). 
This will decrease the inclination angle with our simple method. 
Such an extended structure is clearly recognized in, e.g., CO(3--2) integrated intensity map (\S~4), 
which indeed yields a better agreement of an inclination angle with the above-mentioned galactic-scale one. 
Meanwhile, the Component-1 appears elongated 
along the {\it polar direction} of the H$_2$O maser disk. 
This immediately brings to mind the pc-scale MIR polar elongation \citep{2014A&A...563A..82T}. 
The derived source size ($1.6 \pm 0.3$ pc $\times$ $1.2 \pm 0.1$ pc) and PA ($298\arcdeg \pm 21\arcdeg$) of Component-1 
are roughly consistent with those of the MIR polar elongation ($\sim 1.9$ pc and $\sim 287\arcdeg$, respectively). 
Therefore, we suggest that Component-1 traces the same physical structure as the MIR polar elongation. 
Although we need higher resolution direct imaging 
to robustly reveal this compact structure, 
our finding here will support the actual existence of such elongated structures 
at the hearts of AGNs \citep[e.g.,][]{2016A&A...591A..47L}, 
which created a new challenge to the classic torus paradigm. 

\begin{table}
\begin{center}
\caption{Results of the visibility modeling (Band 7) \label{tbl3}}
\begin{tabular}{ccc}
\tableline\tableline
 & Component-1 & Component-2 \\ 
\tableline
Major axis (mas) & 80 $\pm$ 13 & 1634 $\pm$ 34 \\ 
Minor axis (mas) & 60 $\pm$ 6 & 951 $\pm$ 21 \\ 
Major axis (pc) & 1.6 $\pm$ 0.3 & 33.3 $\pm$ 0.7 \\ 
Minor axis (pc) & 1.2 $\pm$ 0.1 & 19.4 $\pm$ 0.4 \\ 
PA ($\arcdeg$) & 298 $\pm$ 21 & 218 $\pm$ 1.4 \\ 
Integrated flux (mJy) & 20.7 $\pm$ 1.3 & 73.5 $\pm$ 1.6 \\ 
\tableline
\end{tabular}
\tablecomments{The two-dimensional double Gaussian fit shown here was performed with the CASA task UVMULTIFIT. 
Both components have the centroid consistent with the AGN position.}
\end{center}
\end{table}

\section{Gas Distributions and Line Ratios}\label{sec4} 
In this section, we analyze the unprecedented 
high-resolution CO(3--2) and [\ion{C}{1}](1--0) 
emission line data of Circinus and then discuss their spatial distributions. 
Our analysis reveals that the bulk of these emission 
comes from the innermost 10$\arcsec$ region, 
although the missing flux is considerable. 
Here, the CASA task \verb|IMMOMENTS| was used to generate the 0$^{\rm th}$ moment maps without any clipping: 
the emission has been integrated over $V_{\rm LSR} = 200-700$ km s$^{-1}$, 
which is sufficient to cover their full velocity ranges (see also Figure \ref{fig11}). 

\subsection{CO(3--2) map}\label{sec4.1} 
\begin{figure*}[h]
\begin{center}
\includegraphics[scale=0.5]{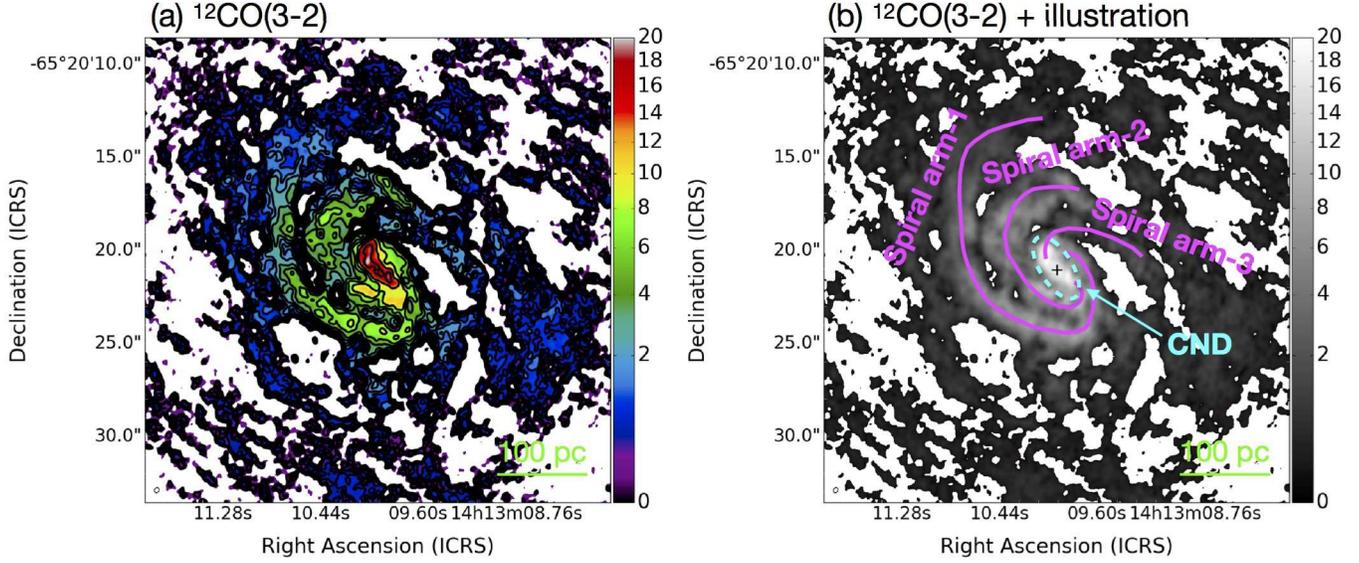}
\caption{
(a) The global integrated intensity map of CO(3--2) in the central 
$25\arcsec \times 25\arcsec$ (or $\sim$ 500 pc $\times$ 500 pc) box of the Circinus galaxy, 
shown in both color scale (Jy beam$^{-1}$ km s$^{-1}$) 
and contours ($5 \times 2^{(n-1)/2}~\sigma$ with $n$ = 1, 2, $\dots$, and 15, 
where 1$\sigma$ = 0.026 Jy beam$^{-1}$ km s$^{-1}$ or 3.8 K km s$^{-1}$). 
The synthesized beam ($0\arcsec.29 \times 0\arcsec.24$) is shown at the bottom left corner as a small white ellipse. 
The primary beam attenuation is uncorrected. 
The AGN position is indicated by a black plus sign. 
(b) The same as (a), but displayed in grayscale. 
The supposed galactic scale major molecular structures 
(i.e., three spiral arms and the CND) are illustrated. 
Note that the southeastern side is the near side of this galaxy. 
}
\label{fig6}
\end{center}
\end{figure*}

\begin{figure*}[h]
\begin{center}
\includegraphics[scale=0.5]{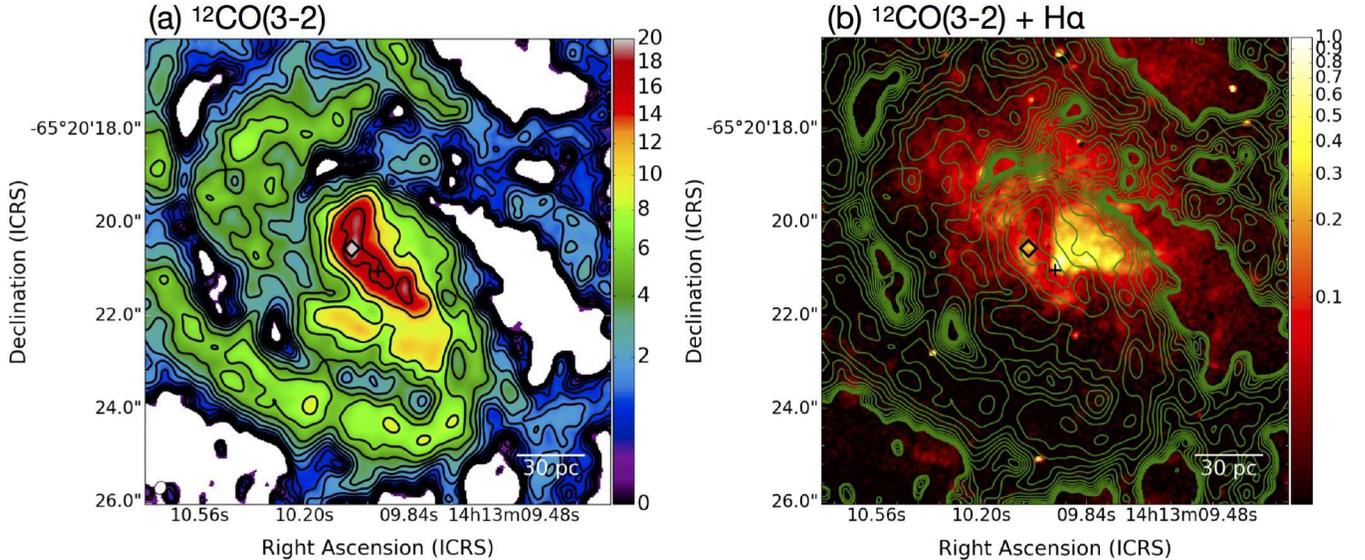}
\caption{
(a) A zoomed-in view of Figure \ref{fig6}a at the central 
$10\arcsec \times 10\arcsec$ (or $\sim 200 \times 200$ pc$^2$) box, 
shown in both color scale (Jy beam$^{-1}$ km s$^{-1}$) 
and contours ($5 \times 2^{(n-1)/2}~\sigma$ with $n$ = 1, 2, $\dots$, and 15, 
where 1$\sigma$ = 0.026 Jy beam$^{-1}$ km s$^{-1}$ or 3.8 K km s$^{-1}$). 
The synthesized beam ($0\arcsec.29 \times 0\arcsec.24$) is 
shown at the bottom left corner as a white ellipse. 
The primary beam attenuation is uncorrected. 
(b) Overlay of the CO(3--2) distribution (contours; levels as in panel (a)) 
on the H$\alpha$ image (color scale in units of counts s$^{-1}$ pixel$^{-1}$). 
In each panel, the central plus sign indicates the AGN location, 
whereas the diamond indicates the peak position of the CO(3--2) integrated intensity. 
}
\label{fig7}
\end{center}
\end{figure*}

Figure \ref{fig6}a shows the CO(3--2) velocity-integrated intensity map of Circinus, 
over the central 25$\arcsec$ (= 500 pc) region. 
The structures seen here are the counterparts to 
the postulated 300 pc-radius ring or disk in \citet{1998MNRAS.300.1119E}. 
After convolving the map to an 18$\arcsec$ aperture and correcting for the primary beam attenuation, 
we found that a roughly $\sim 50\%$ of the total CO(3--2) line flux has been resolved out 
compared to the APEX single dish data \citep{2014A&A...568A.122Z}. 
As for the dust continuum emission, we identify two main structures 
in this circumnuclear region through closer inspection of the map: 
these are illustrated in Figure \ref{fig6}b. 
A zoomed-in view (central 10$\arcsec$ = 200 pc) is shown in Figure \ref{fig7}a. 
We describe each structure below. 

\subsubsection{Spiral arms}\label{sec4.1.1}
Three possible spiral arms (labeled 1, 2, and 3 in Figure \ref{fig6}b) are visible. 
These could be inner extensions of the arms seen in the larger scale CO(1--0) map \citep{2016ApJ...832..142Z}. 
In addition, inter-arm spur or feather structures also appear, 
as is the case in the nearby grand-design spiral galaxy M51 \citep[e.g.,][]{2013ApJ...779...42S}. 
Meanwhile, the gaseous bar at the southeast of the nucleus \citep{2000ApJ...531..219M} cannot be identified: 
the relatively straight-aligned CND and bright knots in the spiral arms 1 and 2 
at the southeastern region of the nucleus (Figure \ref{fig7}a) could have been mistaken as a bar. 

Given the high critical density of CO(3--2), $n_{\rm crit} = 5 \times 10^4$ cm$^{-3}$ \citep[for a temperature of 100 K,][]{1999ApJ...527..795K}, 
one would expect that several bright {\it knots} in the spiral arms (Figure \ref{fig7}a) are the sites of star formation: 
each knot typically has a size of a few pc to $\sim 10$ pc, 
and thus can be classified as a molecular cloud. 
Then, the spatial distributions of the CO(3--2) and the H$\alpha$ (\S~2) are compared in Figure \ref{fig7}b. 
At spiral arm 2 and 3, where the H$\alpha$ emission is weakly visible, 
there were no good spatial coincidences between 
the CO-bright knots/regions and the H$\alpha$-bright regions. 
Therefore, as is naively expected, this H$\alpha$ emission 
only traces unobscured star formation. 
Another finding is that the spiral arm 1 clearly traces 
the H$\alpha$ dark lane at the southeastern side (near side) of this galaxy \citep{2000AJ....120.1325W}. 
If we suppose that typical CO(3--2) emission along the spiral arm 1 
($\sim 4$ Jy beam$^{-1}$ km s$^{-1}$) is thermalized with that of CO(1--0), 
we would expect there to be visual extinction of $A_V \sim 120$ mag, 
by applying the standard Galactic CO conversion factor and the extinction law
\footnote{$X_{\rm CO(1-0)} = 2 \times 10^{20}$ cm$^{-2}$ (K km s$^{-1}$)$^{-1}$ from \citet{2013ARA&A..51..207B} 
and $A_V/N_{\rm H} = 5.3 \times 10^{-22}$ mag cm$^2$ from \citet{2011piim.book.....D}.}. 
This is one order of magnitude higher than that estimated from 
the $R$--$H$ and $H$--$K$ color map analyses \citep{2000ApJ...531..219M}, 
and thus is sufficient to create the dark lane.

\subsubsection{The CND}\label{sec4.1.2}
The spiral arms converge to, or jointly constitute, 
a bright and compact gas concentration at the center, i.e., the CND, 
which appears as an inclined disk. 
A two-dimensional Gaussian fit to the central $3\arcsec \times 3\arcsec$ area 
on the image plane (CASA task \verb|IMFIT|) 
estimates the beam-deconvolved distribution as 
($3\arcsec.62 \pm 0\arcsec.15$) $\times$ ($1\arcsec.66 \pm 0\arcsec.07$) 
or (74 $\pm$ 3) pc $\times$ (34 $\pm$ 1) pc with PA = $212.0\arcdeg \pm 1.9\arcdeg$. 
The derived PA here (and that of the Component-2 found in the Band 7 continuum emission distribution) 
is entirely consistent with those on the galactic $>$ kpc scale, 
e.g., \ion{H}{1} emission and CO(1--0) emission distributions \citep[$\sim 210\arcdeg$, e.g.,][]{1977A&A....55..445F,1998A&A...338..863C}, 
as well as that of the nuclear H$_2$O maser disk 
\citep[209$\arcdeg$ at $r = 0.1$ pc and 236$\arcdeg$ at $r = 0.4$ pc;][]{2003ApJ...590..162G}. 
An inclination angle estimated from the above deconvolution, 
by assuming that the molecular gas is distributed in a circular disk, is $\simeq 63\arcdeg$. 
This angle is again fully consistent with that of the galactic scale \ion{H}{1} gas distribution 
but is smaller than that of the nuclear MIR disk \citep[$> 75\arcdeg$,][]{2014A&A...563A..82T}. 
The global inclination angle we obtained is also smaller than that ($\sim 75\arcdeg$) 
required in \citet{2016ApJ...828L..19W} to reproduce the continuum SED 
of Circinus with the multi-phase dynamic torus model. 
One plausible and simple explanation is that the CND of Circinus is warped 
from the center \citep[i.e., the edge-on disk seen as the H$_2$O maser disk,][]{2003ApJ...590..162G} to the outward edge. 
This is in agreement with the observation that the orientation of 
a large-scale gas/stellar distribution is unrelated 
to that of the radio jet, i.e., the polar direction of a presumed torus 
\citep{1998ApJ...495..189C,1999ApJ...516...97N}. 

We found good spatial coincidence between 
the Band 7 continuum distribution and the CND. 
Thus the CND also delineates the outer boundary 
of the warm dust structure \citep{2016MNRAS.457L..94M}, 
which would collimate the H$\alpha$ emission, as shown in \S~3. 
Note that we suppose that the northwestern part of the spiral arm 2 (see also Figure \ref{fig6}b) 
is at the far side with respect to the H$\alpha$ cone: 
otherwise the cone is obscured by the cold dusty gas. 
The apex of the H$\alpha$ cone barely appears because this CND is inclined. 
Meanwhile, the other side (i.e., southeastern side) of 
the H$\alpha$ cone is completely obscured by the CND, 
although that side is partially visible at longer wavelength emission lines 
such as Pa$\alpha$ \citep{2000ApJ...531..219M,2016MNRAS.457L..94M}. 

Owing to the much higher S/N (Table \ref{tbl2}) than the Band 7 continuum map, 
our CO(3--2) integrated intensity map reveals a wealth of detail in this CND. 
The gas distribution there is not uniform with the off-centered peak position 
(20.8 Jy beam$^{-1}$ km s$^{-1}$; marked as the diamond in Figure \ref{fig7}) 
at $\sim 1\arcsec$ northeast of the nucleus (18.0 Jy beam$^{-1}$ km s$^{-1}$). 
This offset would be unaffected by spatial filtering given 
the much larger maximum recoverable scale ($\sim 7\arcsec$) of our observations (\S~2). 
The spatial offset of the CO peaks with respect to the exact AGN position is consistent with 
the highly inhomogeneous nature of CO brightness distributions 
simulated with our multi-phase dynamic torus model \citep{2018ApJ...852...88W}. 

We estimate the molecular mass of the CND to be $M_{\rm H_2} \simeq 3 \times 10^6~M_\odot$, as follows. 
Here, we define the extent of the CND as that of the above beam-deconvolved source size 
($3\arcsec.62 \times 1\arcsec.66$, PA = 212$\arcdeg$.0). 
We assume that the CO(3--2) emission there (integrated intensity = 805 Jy km s$^{-1}$) 
is thermalized with that of CO(1--0) and use the standard equation \citep{2005ARA&A..43..677S} 
to calculate its line luminosity as $L'_{\rm CO(1-0)} = 3.85 \times 10^6$ K km s$^{-1}$ pc$^2$. 
Then, by applying the canonical CO conversion factor in active environments 
\citep[$\alpha_{\rm CO(1-0)}$ = 0.8 $M_\odot$ (K km s$^{-1}$ pc$^2$)$^{-1}$,][]{1998ApJ...507..615D,2013ARA&A..51..207B}, 
the above $M_{\rm H_2}$ is obtained. 
In the same manner, we also estimate the line-of-sight H$_2$ column density toward the AGN position 
from the CO(3--2) integrated intensity (18.0 Jy beam$^{-1}$ km s$^{-1}$, or 2675 K km s$^{-1}$) 
as $N_{\rm H_2} = 5.4 \times 10^{23}$ cm$^{-2}$. 
Note that we expect this $N_{\rm H_2}$ to remain the lower limit 
of the true line-of-sight column density due to beam dilution. 
For example, given the $\sim 3$ K brightness temperature 
of the Band 7 continuum emission at the AGN position, 
we may expect $\gtrsim$ a few to even $\sim 10\times$ larger $N_{\rm H_2}$ 
if the true physical temperature of the cold dust is, e.g., $\gtrsim 30$ K (with a modest opacity). 
Moreover, if $\alpha_{\rm CO}$ is larger than the above canonical value by a factor of $\sim 4-5$, 
as expected in recent radiative transfer simulations of multiple CO lines \citep{2018ApJ...852...88W},
we will also obtain a correspondingly larger $M_{\rm H_2}$ and $N_{\rm H_2}$. 
Thus, this CND would provide a significant fraction of the Compton-thick material 
toward the Circinus AGN \citep[$N_{\rm H} = (6.6-10) \times 10^{24}$ cm$^{-2}$,][]{2014ApJ...791...81A}. 
This is consistent with the results of our multi-phase dynamic torus model 
applied to Circinus \citep{2016ApJ...828L..19W}, 
which required additional nuclear obscuration by the CND-scale cold ISM 
to the obscuration by the central warm ISM, 
to fully reproduce the SED of this AGN. 
Hence, we suggest that the CND indeed operates as a significant part of the nuclear obscurer.

\subsection{\cone map}\label{sec4.2} 
\begin{figure*}
\begin{center}
\includegraphics[scale=0.5]{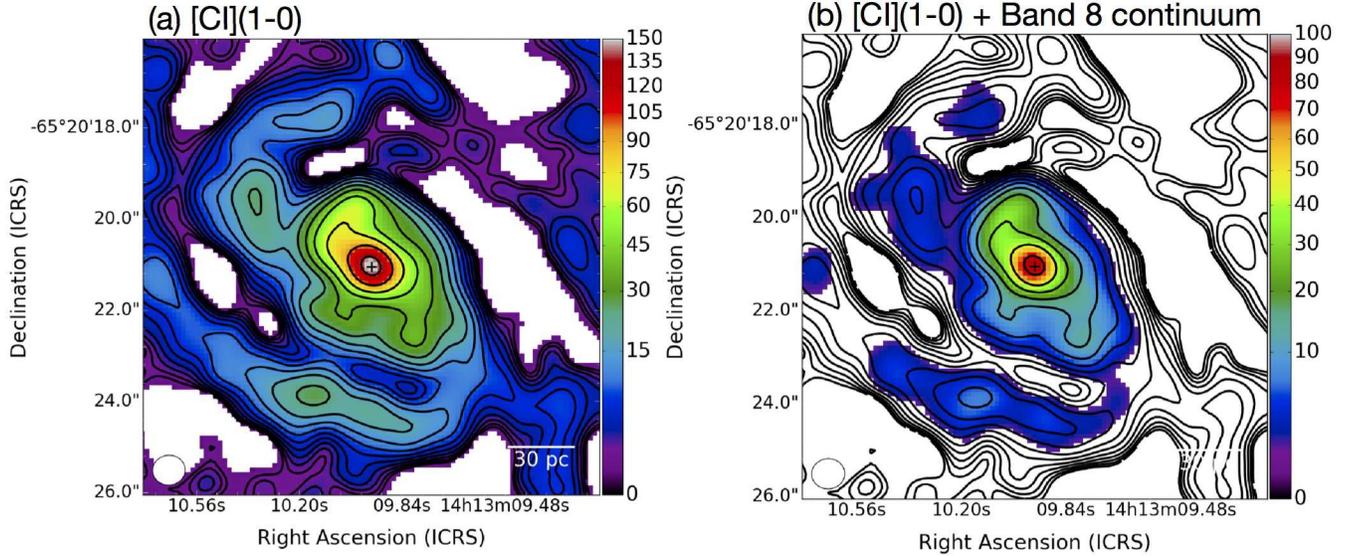}
\caption{
(a) The integrated intensity map of [\ion{C}{1}](1--0) in the central 
$10\arcsec \times 10\arcsec$ (or $\sim$ 200 pc $\times$ 200 pc) box of the Circinus galaxy, 
shown in both color scale (Jy beam$^{-1}$ km s$^{-1}$) 
and contours ($5 \times 2^{(n-1)/2}~\sigma$ with $n$ = 1, 2, $\dots$, and 15, 
where 1$\sigma$ = 0.215 Jy beam$^{-1}$ km s$^{-1}$). 
The synthesized beam ($0\arcsec.71 \times 0\arcsec.66$) is shown at the bottom left corner as a white ellipse. 
The primary beam attenuation is uncorrected here. 
The AGN position is marked by a black plus sign. 
(b) The overlay of the [\ion{C}{1}](1--0) integrated intensity (contours) 
on the Band 8 continuum map (color scale; same as in Figure \ref{fig2}b). 
}
\label{fig8}
\end{center}
\end{figure*}

The \cone line emission is successfully detected both 
at the spiral arms and the CND (Figure \ref{fig8}a), 
with a clear peak at the AGN position (154 Jy beam$^{-1}$ km s$^{-1}$). 
The \cone distribution is consistent with that of the simultaneously 
obtained Band 8 continuum emission (Figure \ref{fig8}b). 
This \cone map shows the global distribution of cold and low-density 
\citep[critical density $n_{\rm crit} = 5 \times 10^2$ cm$^{-3}$,][]{1999ApJ...527..795K} gas in Circinus. 
It is difficult to estimate the missing flux for this \cone observation 
as the field of view of ALMA Band 8 (12$\arcsec$.8) is even smaller 
than the mapped area (18$\arcsec$) with APEX reported in \citet{2014A&A...568A.122Z}. 
However, we found that our observation recovered 45\% of that single dish flux 
after correcting for the primary beam attenuation. 

\begin{figure}
\begin{center}
\includegraphics[scale=0.52]{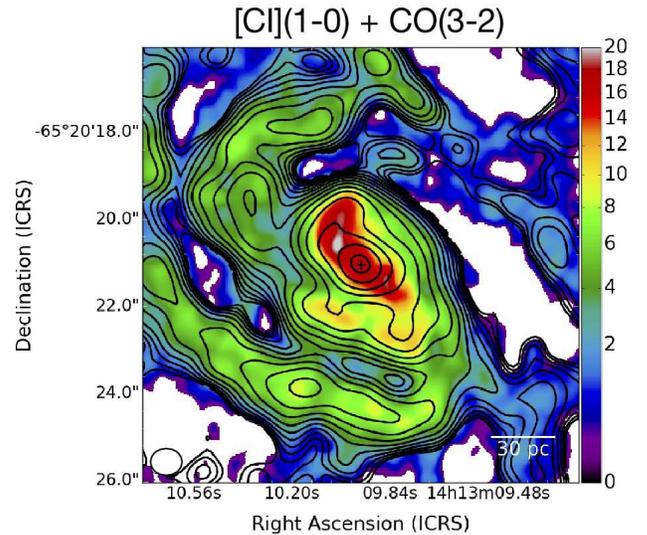}
\caption{
The superposition of the \cone integrated intensity map (contours, as in Figure \ref{fig8}) 
on the CO(3--2) integrated intensity map (color scale, as in Figure \ref{fig7}), 
in the central $10\arcsec \times 10\arcsec$ (200 pc $\times$ 200 pc) of the Circinus galaxy. 
The central plus sign marks the AGN location. 
The synthesized beam of the \cone cube is identified by the bottom left ellipse. 
}
\label{fig9}
\end{center}
\end{figure}

The overall resemblance of the \cone distribution and the CO(3--2) distribution (Figure \ref{fig9}) 
supports the previous argument that \cone essentially traces the same area as low-$J$ CO lines, 
which has previously been reported for Galactic molecular clouds 
\citep[e.g.,][]{1999ApJ...527L..59I,2000ApJ...539L.133P,2005ApJ...623..889O} 
and extragalactic objects \citep{2016A&A...592L...3K,2018arXiv180200486M}, 
and has been found in numerical models \citep[e.g.,][]{2015MNRAS.448.1607G}. 
However, we also found slight spatial inconsistencies in the detailed gas distributions, 
both at the spiral arms and the CND (Figure \ref{fig9}). 
Therefore, the \cone also traces different gas volume from that the CO(3--2) does. 

With regard to the spiral arms, we suggest that the spatial offsets ($\lesssim 10$ pc) 
of the \cone peaks and the CO(3--2) peaks are due primarily to excitation conditions, 
as CO(3--2) requires $\sim 100 \times$ denser gas than [\ion{C}{1}](1--0) to be excited. 
\citet{2018arXiv180200486M} also reported spatial offsets between 
[\ion{C}{1}](1--0)-bright knots and CO(3--2)-bright knots 
in the circumnuclear starburst ring of NGC 613. 
Meanwhile, our particular interest in this paper is 
centered on the [\ion{C}{1}](1--0) to CO(3--2) spatial offset inside the CND. 
The \cone peaks exactly at the AGN location, 
while the CO(3--2) peak is $\sim 1\arcsec$ northeast 
of that position (Figure \ref{fig9}); this is discussed in more detail in \S~4.3. 

We also performed a first-order estimate of $M_{\rm H_2}$ from 
the \cone line luminosity ($L'_{\rm CI(1-0)}$) at the CND, 
defined by the beam-deconvolved source size ($3\arcsec.62 \times 1\arcsec.66$) from the CO(3--2) analyses (\S~4.1). 
The obtained value there is 1.42 $\times$ 10$^6$ K km s$^{-1}$ pc$^2$. 
We first estimate the atomic carbon mass ($M_{\rm CI}$) 
by following \citet{2002ApJS..139..467I} and \citet{2005A&A...429L..25W} as, 
\begin{equation}\label{eq1}
M_{\rm CI} = 5.71 \times 10^{-4} Q(T_{\rm ex}) \frac{1}{3}e^{23.6/T_{\rm ex}}L'_{\rm CI(1-0)} ~[M_\odot], 
\end{equation}
where $Q(T_{\rm ex}) = 1 + 3e^{-T_1/T_{\rm ex}} + 5e^{-T_2/T_{\rm ex}}$ is the \ion{C}{1} partition function, 
$T_1 = 23.6$ K and $T_2 = 62.5$ K are the level energies above the ground state, 
and $T_{\rm ex}$ is the excitation temperature. 
This equation assumes optically thin line emission under the LTE condition. 

Even under the LTE condition, we would need to correct for line opacity 
as there is no guarantee that the \cone line emission in Circinus is optically thin. 
However, we expect that its line opacity is moderate ($\tau_{\rm CI(1-0)} \lesssim 1$) 
given its peak line intensity ($\sim 1144$ mJy beam$^{-1}$ or $\sim 12.3$ K) at the AGN position 
and its likely high $T_{\rm ex}$ such as $\gtrsim 30$ K \citep[typical value observed in high-redshift quasars,][]{2011ApJ...730...18W}. 
If $\tau_{\rm CI(1-0)} \lesssim 1$, then the correction factor for flux attenuation ($\tau_{\rm CI(1-0)}$/$1-\exp(-\tau_{\rm CI(1-0)})$) is $\lesssim 1.6$, 
i.e., the equation (\ref{eq1}) gives a good estimate of $M_{\rm CI}$. 
Then we achieve $M_{\rm CI} \simeq 1770~M_\odot$ within the CND by assuming that $T_{\rm ex} = 30$ K; 
this mass does not change significantly as long as $T_{\rm ex} \gtrsim 30$ K. 
If we adopt a relative \ion{C}{1} abundance with respect to H$_2$ as $8 \times 10^{-5}$, 
which is also the typical value observed in high-redshift quasars \citep{2011ApJ...730...18W}, 
we obtain $M_{\rm H_2}$ of $\simeq 3.7 \times 10^6~M_\odot$ in the CND. 
This is consistent with the CO(3--2)-based $M_{\rm H_2}$ derived in \S~4.1.2. 

\subsection{[\ion{C}{1}](1--0)/CO(3--2) line ratio}\label{sec4.3}
\begin{figure}
\begin{center}
\includegraphics[scale=0.45]{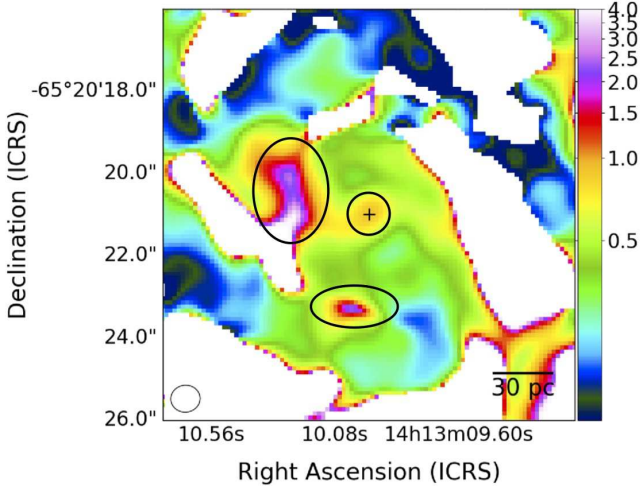}
\caption{
The \cone to CO(3--2) integrated intensity ratio map (brightness temperature scale) 
in the central $10\arcsec \times 10\arcsec$ (200 pc $\times$ 200 pc) of the Circinus galaxy. 
Three regions of high line ratio (black ellipses), one at the AGN position 
and two at the spiral arms, are highlighted for clarity. 
This map is made over the region where 
both the CO(3--2) and the \cone integrated intensity maps show $S/N > 5$. 
Note that the $uv$- and beam size-matched CO(3--2) data is used here. 
}
\label{fig10}
\end{center}
\end{figure}

Here, we discuss the [\ion{C}{1}](1--0)/CO(3--2) integrated intensity ratio 
($\equiv R_{\rm CI/CO}$; brightness temperature scale) at both the CND and the spiral arms 
by comparing them with previous measurements in external galaxies and Galactic star-forming regions. 
We first adjust the $uv$ range and the beam size of the CO(3--2) data 
to those of the \cone data (25--470 k$\lambda$, $0\arcsec.71 \times 0\arcsec.66$ with PA = 96$\arcdeg$). 
The resultant CO(3--2) cube has a 1$\sigma$ sensitivity of 0.54 mJy beam$^{-1}$ at a velocity resolution of 10 km s$^{-1}$. 

\begin{figure}
\begin{center}
\includegraphics[scale=0.47]{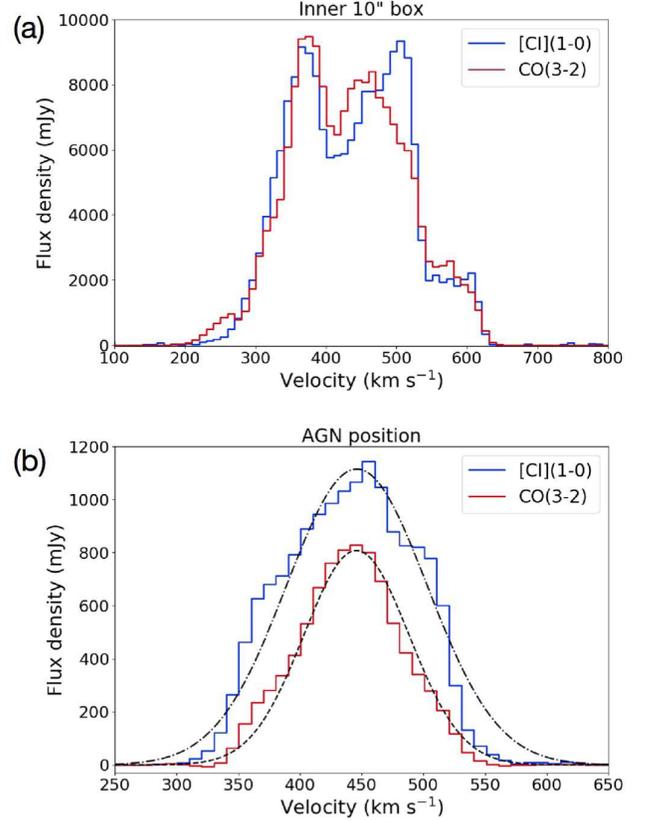}
\caption{
Line profiles of the \cone and the CO(3--2) measured at (a) the inner $10\arcsec \times 10\arcsec$ box 
and (b) the AGN location (with the single $0\arcsec.71 \times 0\arcsec.66$ beam) of the Circinus galaxy. 
The $uv$ range and the beam size of the CO(3--2) data are matched to those of the \cone data. 
The multiple horn-line profile in (a) is due to the spiral arms. 
Results of single Gaussian fits to the CO(3--2) and \cone spectra at the AGN position 
(indicated by the dashed line and dot-dashed line, respectively) are summarized in Table \ref{tbl5}. 
}
\label{fig11}
\end{center}
\end{figure}

Figure \ref{fig10} shows the $R_{\rm CI/CO}$ in the central region of Circinus. 
The typical value of $R_{\rm CI/CO}$ at the spiral arms is $\sim 0.3$, 
although there are two spots with high ratios; we expect that 
the local gas densities there are too low to efficiently excite the CO(3--2) line. 
On the other hand, it is notable that the $R_{\rm CI/CO}$ 
at the AGN position is $\sim 3$ times higher ($\sim 0.9$) than the spiral arms. 
Given the much higher $n_{\rm crit}$ of CO(3--2) than that of [\ion{C}{1}](1--0), 
this ratio would be hard to explain by simple gas excitation as it would 
be naively expected that the gas density profile is centrally peaked. 
The same situation is also demonstrated in the line spectra (Figure \ref{fig11}): 
while the \cone and the CO(3--2) show totally comparable line flux densities 
when measured at the central 10$\arcsec$ box (i.e., CND + spiral arms), 
the \cone emission clearly becomes brighter than the CO(3--2) emission at the AGN position. 
The relevant $R_{\rm CI/CO}$ measured at several spatial scales are listed in Table \ref{tbl4}; 
the high $R_{\rm CI/CO}$ at the AGN position is uncommon in star-forming galaxies, 
which could be explained by an efficient CO dissociation due to hard X-ray irradiation from the AGN 
\citep[XDR,][]{1996ApJ...466..561M,2005A&A...436..397M}. 

Indeed, based on the one-zone XDR model of \citet{1996ApJ...466..561M}, 
we can estimate the X-ray energy deposition rate per particle ($H_X/n$) with 
\begin{equation}
H_X \sim 7 \times 10^{-22} L_{44} r^{-2}_2 N^{-1}_{22} ~{\rm erg~s^{-1}}, 
\end{equation}
where $L_{44}$ is the 1--100 keV X-ray luminosity in units of 10$^{44}$ erg s$^{-1}$, 
$r_2$ is the distance from the AGN to the point of interest in units of 100 pc, 
and $N_{22}$ is the X-ray attenuating column density in units of 10$^{22}$ cm$^{-2}$, respectively. 
Here, we adopted $L_{44} = 0.13$ (with a photon index = 1.31 and cut-off energy = 160 keV) from \citet{2014ApJ...791...81A}, 
as well as $n_{\rm H_2} = 10^5$ cm$^{-3}$ as a typical value in the molecular part 
of CNDs of Seyfert galaxies \citep[e.g.,][]{2013PASJ...65..100I,2014A&A...570A..28V}. 
Then, we estimated $H_X/n$ at $r = 7$ pc from the center 
(i.e., the half-major axis of the \cone synthesized beam) as $\log(H_X/n) = -27.3$. 
According to the models in \citet{1996ApJ...466..561M}, 
this is almost exactly the rate at which the fractional abundance 
of \ion{C}{1} ($X_{\rm CI}$) becomes equal to that of CO ($X_{\rm CO}$). 
At the regions where $\log(H_X/n) > -27.3$, we can expect $X_{\rm CI}/X_{\rm CO} > 1$. 
Thus, we suggest that the molecular dissociation due to X-ray irradiation is significant 
at the close vicinity of the Circinus AGN, which could cause the high $R_{\rm CI/CO}$ there. 

\begin{table}
\begin{center}
\caption{The [\ion{C}{1}](1--0)/CO(3--2) line ratio \label{tbl4}}
\begin{tabular}{cccc}
\tableline\tableline
 & AGN & CND & 10$\arcsec$ box \\ 
\tableline
Jy km s$^{-1}$ & 1.80 $\pm$ 0.25 & 1.03 $\pm$ 0.15 & 0.97 $\pm$ 0.14 \\
K km s$^{-1}$ & 0.89 $\pm$ 0.13 & 0.51 $\pm$ 0.07 & 0.48 $\pm$ 0.07 \\
\tableline
\end{tabular}
\tablecomments{$``$AGN$"$ indicates that the ratio is measured at the exact AGN location 
with the single \cone synthesized beam ($0\arcsec.71 \times 0\arcsec.66$). 
$``$CND$"$ refers to the $3\arcsec.62 \times 1\arcsec.66$ (PA = 212$\arcdeg$) 
region defined in the CO(3--2) map (\S~4.1.2). 
The 10$\arcsec$ box contains both the CND and the spiral arms. 
The 10\% absolute flux uncertainties are included here.}
\end{center}
\end{table}

\begin{table}
\begin{center}
\caption{Results of the single Gaussian fit \label{tbl5}}
\begin{tabular}{cccc}
\tableline\tableline
 & Peak & Centroid & FWHM \\ 
 & (mJy beam$^{-1}$) & (km s$^{-1}$) & (km s$^{-1}$) \\
\tableline
CO(3--2) & 807.9 $\pm$ 0.9 & 445.8 $\pm$ 0.1 & 100.3 $\pm$ 0.1 \\
\cone & 1115.1 $\pm$ 4.6 & 446.1 $\pm$ 0.1 & 134.3 $\pm$ 0.2 \\
\tableline
\end{tabular}
\tablecomments{These fits were performed for the spectra at the AGN position, 
measured with the single \cone synthesized beam (Figure \ref{fig11}b).} 
\end{center}
\end{table}

However, the actual three-dimensional (3D) geometrical structure 
of the XDR requires careful consideration: 
it does not necessarily form a simple one-zone structure. 
Indeed, single Gaussian fits to the observed CO(3--2) 
and \cone spectra measured at the AGN position (Figure \ref{fig11}b and Table \ref{tbl5}) 
revealed that the \cone spectrum has a wider full width at half maximum (FWHM) than the CO(3--2) spectrum, 
as well as deviated components from the single Gaussian profile at about $\pm$70 km s$^{-1}$ 
to the systemic velocity (446 km s$^{-1}$). 
Both of these imply that the \cone emission also traces a different 
gas volume than the CO(3--2) emission. 
This is consistent with the multi-phase dynamic torus model (\S~5 and 6), 
where the complex interplay of gas chemistry and dynamics indeed occurs. 
Furthermore, there could be several other possibilities that can cause high $R_{\rm CI/CO}$, 
including severe self-absorption in the CO(3--2) line and very peculiar gas excitation condition. 
Observations of multiple transition lines and multiple species 
including isotopologues are required to better understand the origin of this enhancement. 

\begin{figure}
\begin{center}
\includegraphics[scale=0.28]{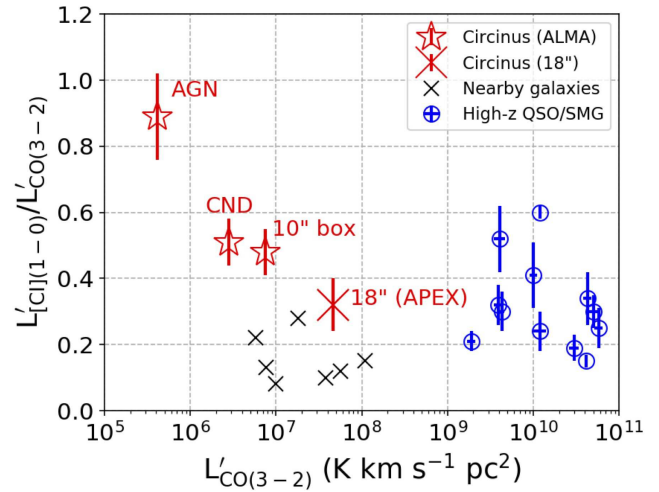}
\caption{
[\ion{C}{1}](1--0)/CO(3--2) line ratio as a function of CO(3--2) line luminosity in units of the brightness temperature. 
The red star indicates the ratios in the Circinus galaxy, measured with three apertures in our ALMA data, 
while the large red cross is the ratio measured with APEX 
\citep[18$\arcsec$ = 370 pc aperture,][]{2014A&A...568A.122Z}. 
Also shown are high-redshift submm galaxies and quasar-host galaxies 
\citep[$\gtrsim 30$ kpc apertures with blue circles,][]{2011ApJ...730...18W}, 
as well as single-dish measurements of nearby galaxies 
\citep[][aperture sizes are several hundreds to $\sim 1$ kpc]{2001A&A...371..433I,2003A&A...404..495I,
2006A&A...445..907I,2004A&A...427...45B}. 
}
\label{fig12}
\end{center}
\end{figure}

Finally, we compare the $R_{\rm CI/CO}$ 
of Circinus measured with three apertures 
with those of other galaxies \citep[compiled from][]{2014A&A...568A.122Z} in Figure \ref{fig12}. 
The comparison data consist of high-z quasar-host galaxies 
and submm galaxies \citep{2011ApJ...730...18W}, 
as well as nearby galaxies including, NGC 6946 and M83 \citep{2001A&A...371..433I}, 
IC 342 \citep{2003A&A...404..495I}, 
Henize 2-10 and NGC 253 \citep{2004A&A...427...45B}, 
and M51 \citep{2006A&A...445..907I}. 
As stated in \citet{2014A&A...568A.122Z}, the $R_{\rm CI/CO}$ 
of the nearby star-forming galaxies are $\sim 0.1$ to 0.2, 
while two AGN-host galaxies M51 and Circinus (18$\arcsec$ aperture) 
show slightly higher values of $\sim 0.3$, 
which already implies a sort of AGN influence on the line ratio. 
The averaged $R_{\rm CI/CO}$ of the high-z objects is 0.32 $\pm$ 0.13 \citep{2011ApJ...730...18W}. 
As these are essentially starburst galaxies \citep[e.g.,][]{2014PhR...541...45C}, 
one plausible explanation for the higher $R_{\rm CI/CO}$ 
compared to nearby galaxies is that their mean interstellar radiation field over the entire galaxy scale 
is as high as those of the central $\sim 1$ kpc of the nearby AGNs, 
which would also lead to efficient CO dissociation. 
Compared to those nearby and high-z samples, 
the increasing trend of $R_{\rm CI/CO}$ toward the center of Circinus, 
which would imply the XDR chemistry, is remarkable.

\section{Molecular and Atomic Gas Dynamics}\label{sec5}
\begin{figure*}
\begin{center}
\includegraphics[scale=0.75]{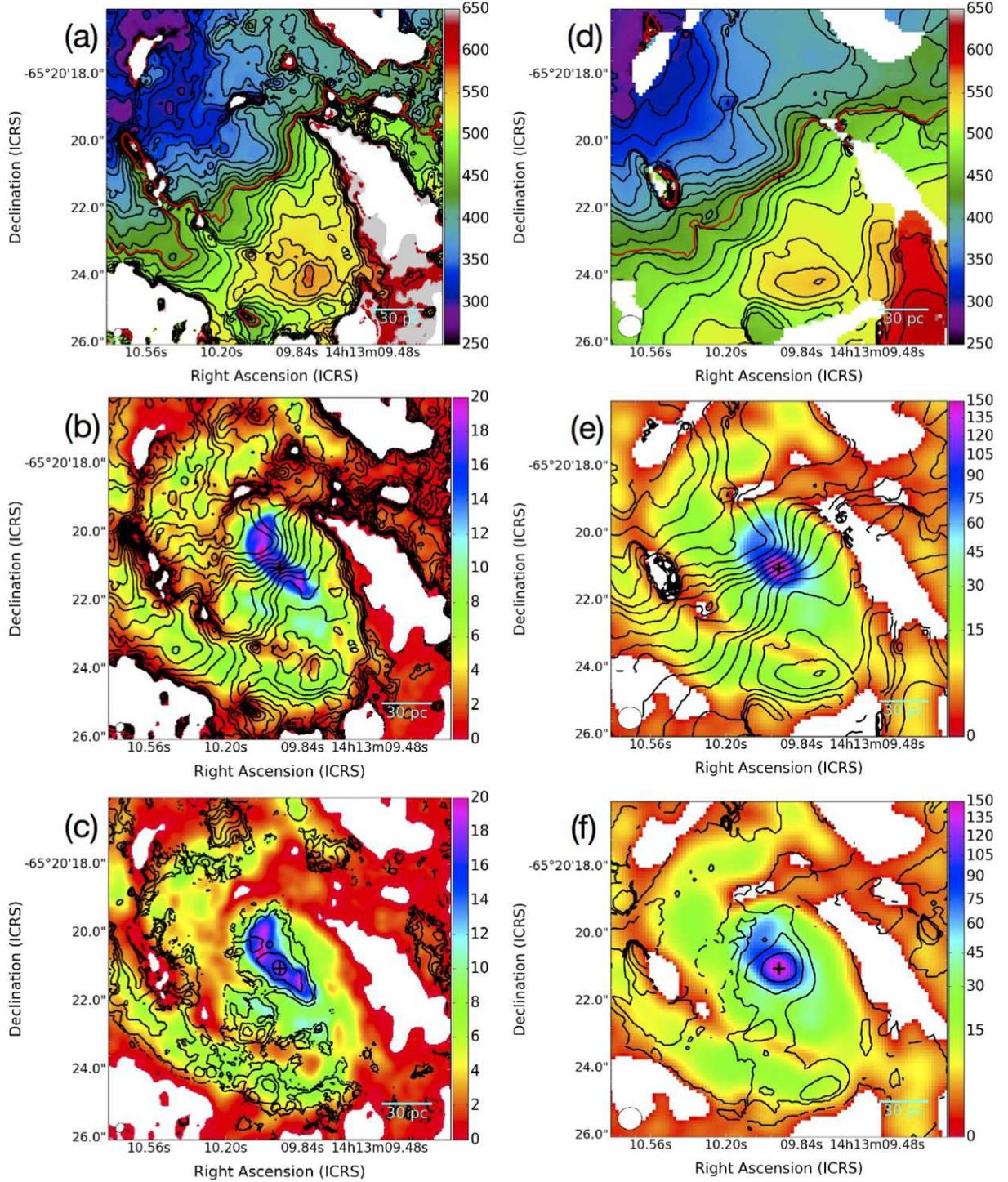}
\caption{
(a) Intensity-weighted mean velocity map of the CO(3--2) 
in the central $10\arcsec \times 10\arcsec$ (200 pc $\times$ 200 pc) box of the Circinus galaxy. 
Contours indicate the line-of-sight velocity of 300--600 km s$^{-1}$ in steps of 15 km s$^{-1}$. 
(b) The same contours as in (a), but superposed on the CO(3--2) integrated intensity map. 
(c) Intensity-weighted velocity dispersion (Gaussian sigma) map of the CO(3--2) 
shown by contours (10--60 km s$^{-1}$ in steps of 10 km s$^{-1}$), 
superposed on the CO(3--2) integrated intensity map. 
(d)(e)(f) Similar images to the left column, but for the cases of the \cone are shown. 
The systemic velocity (446 km s$^{-1}$) estimated from our 
single Gaussian fit (Figure \ref{fig11}) 
is highlighted by the red line in both (a) and (d). 
The integrated intensities are shown in units of Jy beam$^{-1}$ km s$^{-1}$ 
and the central plus sign marks the AGN position. 
}
\label{fig13}
\end{center}
\end{figure*}

In this Section, we model the dynamics 
of the cold and dense molecular gas probed by the CO(3--2) line 
and that of the more diffuse atomic gas probed by the \cone line. 
Our aim is to investigate their difference, 
particularly at the CND scale or inside, 
which will be key to understanding the geometrical structure 
of the circumnuclear obscuring material around the AGN. 

\subsection{Global patterns}\label{sec5.1}
First, we show the global gas dynamics (intensity-weighted line-of-sight velocity and velocity dispersion) 
probed by both the CO(3--2) and the \cone emission lines in Figure \ref{fig13}. 
These maps were made with the CASA task \verb|IMMOMENTS| 
with 10$\sigma$ clipping to avoid noisy pixels. 
The patterns traced by the two lines are reasonably consistent with each other, 
although we would need to beam-deconvolve them to achieve 
the actual gas dynamics (i.e., intrinsic rotation and dispersion; \S~5.2). 
The gas motion is clearly dominated by the galactic rotation 
with an overall northeast--southwest orientation (Figure \ref{fig13}a,d), 
which is consistent with previous studies at larger spatial scales 
\citep[e.g.,][]{2008MNRAS.389...63C,2016ApJ...832..142Z}. 
On the other hand, it is also evident in the spiral arms 
that streaming motions are superposed on the rotation pattern (Figure \ref{fig13}b,e). 
These streaming motions, as well as spatial structures in each velocity channel (see Appendix), 
may help the reader to recognize the likely three spiral arms 
that we have postulated and illustrated in Figure \ref{fig6}b.

\subsection{Decomposition with tilted rings}\label{sec5.2}
To extract basic beam-deconvolved dynamical information, 
particularly rotational velocity ($V_{\rm rot}$) and dispersion ($\sigma_{\rm disp}$), 
we fitted concentric titled rings to the observed velocity structures 
with the $^{\rm 3D}$Barolo code \citep{2015MNRAS.451.3021D}. 
The main parameters were dynamical center, systemic velocity ($V_{\rm sys}$), 
$V_{\rm rot}$, $\sigma_{\rm disp}$, galactic inclination ($i$), and PA, 
all of which can be varied in each ring. 
Here we fixed the dynamical center to the AGN position 
and the $V_{\rm sys}$ to 446 km s$^{-1}$ based on 
the single Gaussian fits to the observed nuclear spectra (Table \ref{tbl5}); 
$V_{\rm rot}$, $\sigma_{\rm disp}$, $i$, and PA were thus the major free parameters 
and were evaluated by minimizing $|$model$-$observed data$|$. 
Given the significantly different spatial resolution between the CO(3--2) cube and the \cone cube, 
we first present the results using the full-resolution CO(3--2) cube in \S~5.2.1 
to describe the details of the gas dynamics as much as possible, 
and then present those with the $uv$- and beam-matched CO(3--2) and \cone cubes in \S~5.2.2. 

\subsubsection{The CO(3--2) full-resolution data}
\begin{figure}
\begin{center}
\includegraphics[scale=0.53]{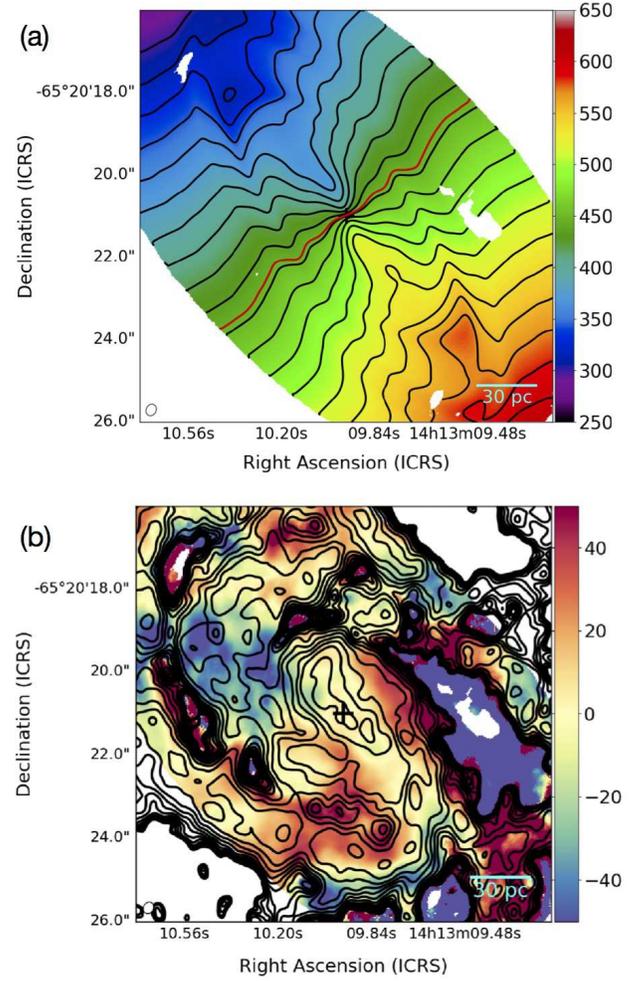}
\caption{
(a) Model velocity field (moment 1) of the Circinus galaxy. 
Contours are the same as in Figure \ref{fig13}(a). 
(b) Residual velocity image after subtracting the model from the observed data (color scale). 
Overlaid contours indicate the velocity-integrated CO(3--2) intensity as shown in Figure \ref{fig7}(a). 
The central plus sign marks the AGN location. 
}
\label{fig14}
\end{center}
\end{figure}

The full-resolution CO(3--2) cube ($\theta = 0\arcsec.29 \times 0\arcsec.24$, $dV = 10$ km s$^{-1}$) was used. 
We modeled 50 concentric rings with a separation of $\Delta r = 0\arcsec.15$, 
which is roughly half the size of the major axis of the synthesized beam. 
The modeled mean velocity field (moment 1) and the residual image after subtracting this model 
from the observed data are displayed in Figure \ref{fig14}a and \ref{fig14}b, respectively. 
The residuals at the spiral arms are $\sim \pm$40 km s$^{-1}$ at the arms 1 and 2, 
which are attributed to streaming motions. 
As the southeastern part is the near side of this galaxy \citep{2000AJ....120.1325W}, 
these motions are likely inflows toward the vicinity of this AGN. 
Meanwhile, the residual is only a few km s$^{-1}$ at the AGN position, 
which is consistent with our torus model tuned for Circinus \citep{2016ApJ...828L..19W}, 
where the inflow velocity through the dense mid-plane of the CND is as slow as $\lesssim 10$ km s$^{-1}$ 
\citep[see also Figure 2 of][]{2012ApJ...758...66W}. 

\begin{figure}
\begin{center}
\includegraphics[scale=0.45]{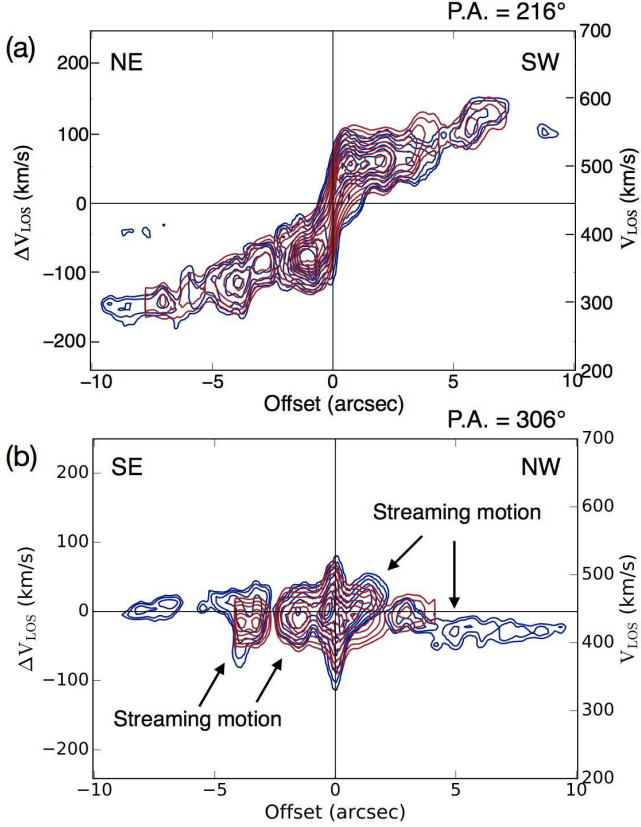}
\caption{
(a) The observed position-velocity diagram (PVD) of the CO(3--2) line 
along the major axis (PA = 216$\arcdeg$) shown as the blue contours. 
The overlaid red contours indicate the PVD produced by our tilted ring model. 
Both contours are plotted at 25, 50, 75, 100, 200, $\cdots$, and 700$\sigma$, 
where 1$\sigma$ = 0.37 mJy beam$^{-1}$. 
The full-resolution data ($\theta = 0\arcsec.29 \times 0\arcsec.24$) is used. 
(b) The PVD along the minor axis (PA = 306$\arcdeg$). 
The same contours as in the panel-(a) are shown. 
Due to the inclined geometry of the concentric rings, 
no model component exists at an offset $\gtrsim 5\arcsec$ (see also Figure \ref{fig14}a). 
}
\label{fig15}
\end{center}
\end{figure}

Figure \ref{fig15}a and \ref{fig15}b presents the modeled 
position-velocity diagrams (PVDs) along the global 
kinematic major and minor axes, respectively, overlaid on the observed PVDs. 
The overall structures are well reproduced by a combination 
of gas rotation and dispersion, although streaming motions are evident 
as well along the minor axis (offset $\sim 2\arcsec$ and $\sim 4\arcsec$), 
which positionally correspond to the spiral arms. 

\begin{figure*}
\begin{center}
\includegraphics[scale=0.54]{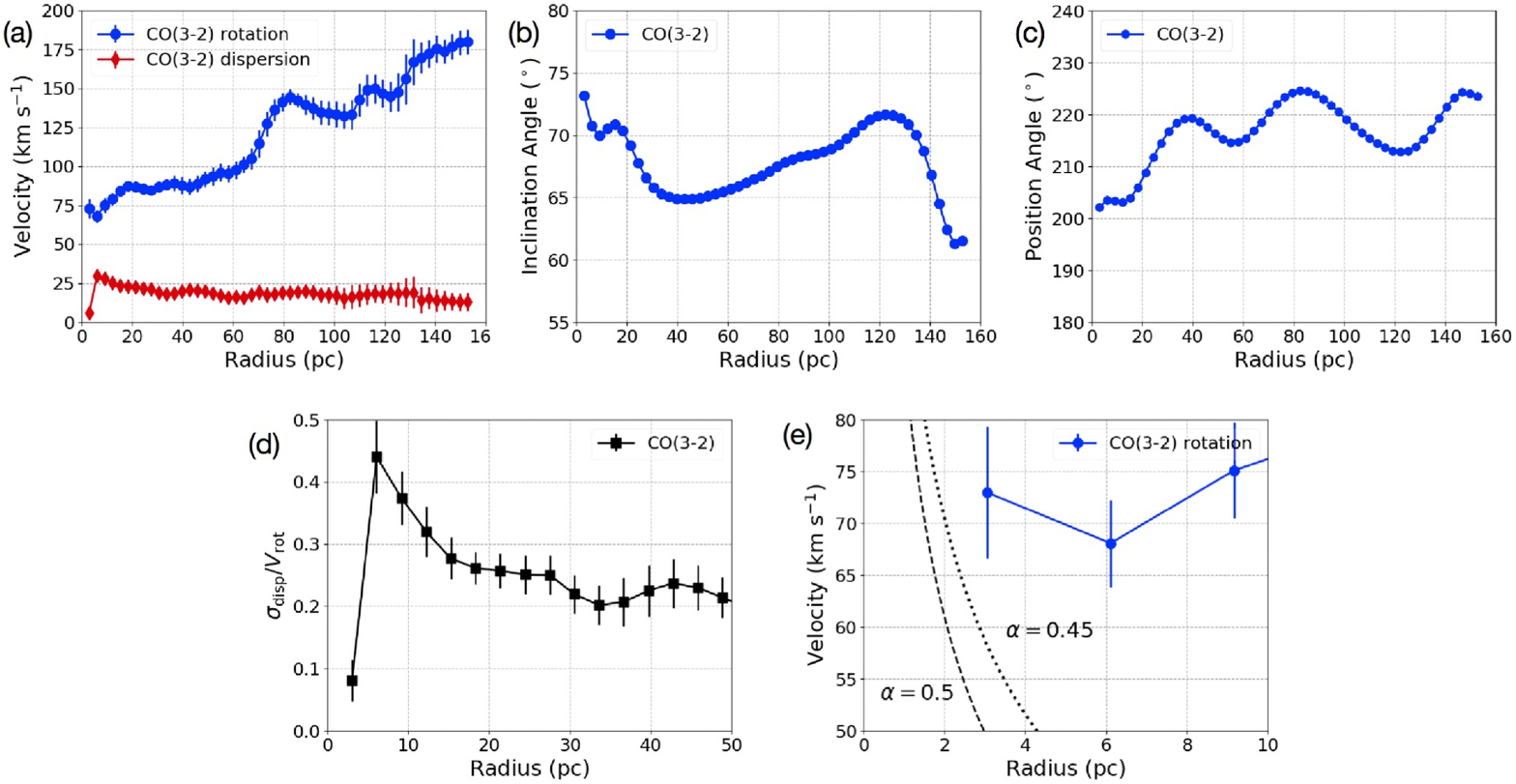}
\caption{
Radial profiles of (a) rotational velocity ($V_{\rm rot}$; blue circles) 
and velocity dispersion ($\sigma_{\rm disp}$; red diamonds), 
(b) inclination angle, and (c) position angle, 
estimated from our tilted ring model of the CO(3--2) velocity field of the Circinus galaxy. 
(d) The radial profile of the $\sigma_{\rm disp}$/$V_{\rm rot}$ ratio. 
(e) A zoomed-in view of the $V_{\rm rot}$ at the innermost 10 pc region. 
A turnover trend of the $V_{\rm rot}$ is suggested at $r \lesssim 5$ pc. 
Also plotted are the (quasi-) Keplerian motion ($V_{\rm rot} \propto r^{-\alpha}$) 
with $\alpha = 0.5$ (dashed) and $\alpha = 0.45$ (dotted), 
inferred from the 22 GHz H$_2$O maser observations \citep{2003ApJ...590..162G}. 
The disk orbital velocity of 260 km s$^{-1}$ at the innermost 0.11 pc 
of that maser disk is used for scaling. 
}
\label{fig16}
\end{center}
\end{figure*}

Figure \ref{fig16}a, b, and c shows the radial profiles of the decomposed 
$V_{\rm rot}$ and $\sigma_{\rm disp}$, $i$, and PA of our model, respectively. 
We found that variation in the $i$ and the PA is small (within $\sim 15\arcdeg$) 
around $i \sim 65\arcdeg$ and PA $\sim 216\arcdeg$. 
These values are consistent with those on the galactic scale \citep[e.g.,][]{1977A&A....55..445F,1998MNRAS.300.1119E}. 
Nevertheless, we found a slight increasing trend in the $i$ toward the AGN position. 
The inferred $i \gtrsim 70\arcdeg$ at $r \lesssim 10$ pc would be consistent with the $i > 75\arcdeg$ 
estimated for the 1 pc scale nuclear MIR disk \citep{2014A&A...563A..82T}, 
as well as with the $i \sim 75\arcdeg$ required to reproduce 
the nuclear IR SED of this AGN based on our torus model \citep{2016ApJ...828L..19W}. 
This increasing $i$ will eventually reach $\sim 90\arcdeg$ as Circinus hosts 
the 22 GHz H$_2$O maser disk at the center \citep{2003ApJ...590..162G}. 

The radial profile of the $\sigma_{\rm disp}$/$V_{\rm rot}$ ratio is shown in Figure \ref{fig16}d. 
This ratio can be considered as a scale height ($H$) 
to scale radius ($R$) ratio ($H/R$, i.e., the aspect ratio of a disk), 
under the hydrostatic equilibrium condition. 
From this panel, one can see that the $\sigma_{\rm disp}$/$V_{\rm rot}$ 
at $r \gtrsim 15$ pc is $\sim 0.25$, 
while the ratio at the central $r \lesssim 10$ pc increases to $\sim 0.4$ 
(except for the innermost $r = 3$ pc ring), 
i.e., the dense molecular disk becomes geometrically thicker. 
Within the qualitative framework of our model, this moderate 
thickness of the dense gas disk at the central $\lesssim$ 10 pc 
can be interpreted such that the dense gas and dust are puffed up 
due to the turbulence induced by the AGN-driven failed winds 
\citep[][see Figure \ref{fig1}]{2012ApJ...758...66W,2016ApJ...828L..19W}. 

It is particularly noteworthy that the innermost $r = 3$ pc ring 
shows a very low $\sigma_{\rm disp}/V_{\rm rot} \sim 0.1$, which is not observed at the other radii. 
The implied very thin disk geometry, as well as the high $i$, would suggest 
that {\it the geometrically thin Keplerian disk has finally started to be captured} 
(i.e., the gas motion is governed by the gravity of the central SMBH) with this cold molecular emission line. 
Indeed, in the zoomed-in view of the radial profile of $V_{\rm rot}$ (Figure \ref{fig16}e), 
we found a turnover trend of $V_{\rm rot}$ at $r = 3$ pc. 
Although the $V_{\rm rot}$ at that radius remains comparable to those at larger radii within uncertainties, 
it is close to the extension of the nuclear Keplerian rotation curve 
around the central SMBH ($M_{\rm BH} = 1.7 \times 10^6~M_\odot$), 
traced by the VLBI 22 GHz H$_2$O maser observations \citep{2003ApJ...590..162G}. 
Although it is difficult at this moment to achieve a firmer conclusion 
given the large errors associated with the inner rings, 
we can observe a clear difference in $V_{\rm rot}$ between 
e.g., $r = 1$ pc and $r = 3$ pc if the CO gas dynamics 
genuinely follow the Keplerian pattern. 
This can be tested with higher-resolution ALMA observations.

\subsubsection{The uv- and beam-matched data}
\begin{figure}
\begin{center}
\includegraphics[scale=0.5]{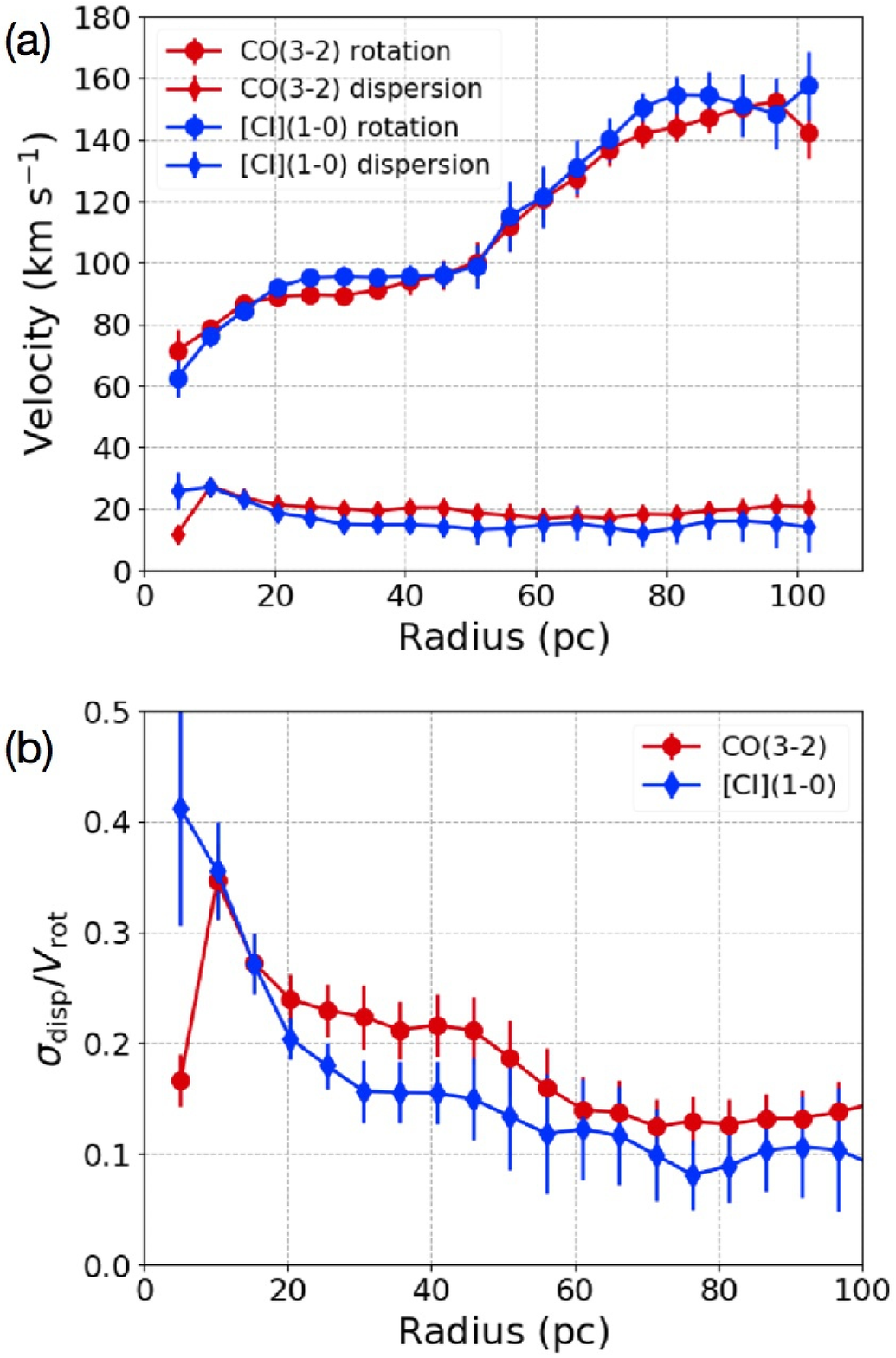}
\caption{
(a) Radial profiles of the rotational velocity ($V_{\rm rot}$; circles) 
and the velocity dispersion ($\sigma_{\rm disp}$; diamonds) 
derived from the CO(3--2) (red) and the \cone (blue) emission lines, 
in the central $r \lesssim100$ pc region of the Circinus galaxy. 
(b) Radial profiles of the $\sigma_{\rm disp}/V_{\rm rot}$ ratios 
traced by the CO(3--2) and the \cone lines, at the same region as in (a). 
The $uv$ ranges and the beam sizes are matched between the CO(3--2) cube and the \cone cube here. 
}
\label{fig17}
\end{center}
\end{figure}

Next, we model the gas dynamics using the \cone cube 
and the $uv$- and beam-matched CO(3--2) cube 
(1$\sigma$ = 0.54 mJy beam$^{-1}$ at $dV$ = 10 km s$^{-1}$). 
With this treatment, we expect to reduce the potential systematic effects 
on the resultant dynamical properties stemming 
from the unmatched $uv$ ranges and beam sizes of the input data. 
Given the larger angular resolution ($0\arcsec.71 \times 0\arcsec.66$), 
we modeled only 20 concentric rings with $\Delta r = 0\arcsec.25$ 
and repeated the procedure in \S~5.2.1. 

Figure \ref{fig17}a displays the resultant radial profiles of 
$V_{\rm rot}$ and $\sigma_{\rm disp}$ traced by the CO(3--2) and the \cone emission lines. 
We found that both the $V_{\rm rot}$ and the $\sigma_{\rm disp}$ of the CO(3--2) derived here 
are consistent with those from the full resolution CO(3--2) cube in \S~5.2.1. 
Thus, the $\sigma_{\rm disp}/V_{\rm rot}$ profile of the CO(3--2) 
shown in Figure \ref{fig17}b is essentially the same as that in Figure \ref{fig16}d. 
Furthermore, the $V_{\rm rot}$ and the $\sigma_{\rm disp}$ 
of the \cone are consistent with those of the CO(3--2) derived here within uncertainties 
at most radii, except for the $\sigma_{\rm disp}$ at the innermost one ($0\arcsec.25 = 5$ pc). 
Considering our model, the similarity in the $\sigma_{\rm disp}/V_{\rm rot}$ ratios 
at $r \sim 10-20$ pc between the two tracers (Figure \ref{fig17}b) 
in turn suggests that the turbulence induced by the failed winds, 
which can act both on the atomic and molecular gas, 
plays a major role in determining the disk geometry there, 
i.e., the contribution from the outflows preferentially seen at ionized/atomic gas 
\citep{2016ApJ...828L..19W} would not be very significant. 
The commonly geometrically thin nature at $r \gtrsim 20$ pc, 
where the radiation-driven fountain will not work well for the case of Circinus \citep{2016ApJ...828L..19W}, 
is also reasonable because SN feedback (an efficient mechanism to puff up the disk) 
should be weak given the low SFR \citep[$\sim 0.1~M_\odot$ yr$^{-1}$;][]{2014ApJ...780...86E}. 

\begin{figure}
\begin{center}
\includegraphics[scale=0.22]{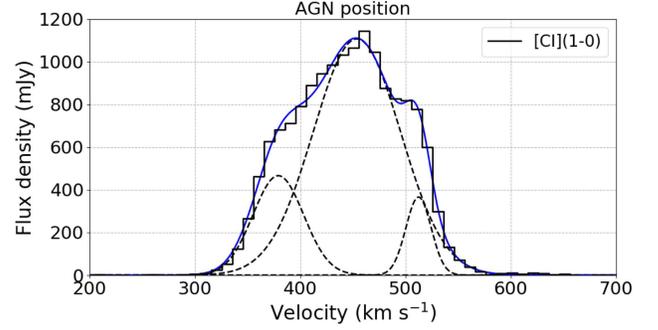}
\caption{
Line profile of the \cone emission measured at the AGN position 
(with the single $0\arcsec.71 \times 0\arcsec.66$ beam) of the Circinus galaxy. 
A three-component Gaussian fit was performed; 
each component is indicated by the black dashed lines, 
and the total model is shown by the blue solid line. 
The derived line properties are listed in Table \ref{tbl6}. 
}
\label{fig18}
\end{center}
\end{figure}

Meanwhile, the discrepancy of the $\sigma_{\rm disp}/V_{\rm rot}$ ratios 
at $r \lesssim 10$ pc ($\sim 0.15$ in the CO(3--2) and $\sim 0.4$ in the [\ion{C}{1}](1--0)) is more evident, 
which indicates the existence of multi-phase dynamical structures around the AGN. 
Note that it is difficult to distinguish coherent outflow motion 
that can widen the line profile from isotropic turbulent motion due to failed winds, 
based simply on the observed $\sigma_{\rm disp}$ that mixes these motions. 
However, as the outflows are preferentially observed 
at ionized and/or atomic gas at the very nuclear region in our model \citep{2016ApJ...828L..19W}, 
we expect that such outflows are the prime source of the geometrically thicker 
nature of the \ion{C}{1} disk than the CO disk at this very central region. 
Indeed, we found that adding two more components (i.e., triple Gaussians) 
can fit the observed nuclear \cone spectrum more faithfully 
as shown in Figure \ref{fig18} (see also Table \ref{tbl6}). 
The central Gaussian component has 
almost the same centroid velocity and FWHM as the CO(3--2) spectrum (Table \ref{tbl5}), 
suggesting that this traces the same region as the CO(3--2) line (the rotating disk mid-plane in our model). 
We also found that the blueshifted and redshifted components 
appear almost {\it symmetrically} with respect to the central component, 
with a velocity offset of $\sim 75$ km s$^{-1}$ (blue) and $\sim 60$ km s$^{-1}$ (red), respectively; 
this is suggestive of coherent atomic gas motion around the AGN. 
{\it Therefore, these results support the view that the diffuse atomic gas is distributed in a geometrically thicker volume 
than the dense molecular gas around the AGN due to atomic outflows}, 
as expected in our multi-phase dynamic torus model. 
As this argument is rather qualitative, however, 
we perform a more quantitative comparison between 
the observed \cone and CO(3--2) line properties 
and our model predictions in the next section.

\begin{table}
\begin{center}
\caption{Results of the three-component Gaussian fit to the observed \cone spectrum \label{tbl6}}
\begin{tabular}{cccc}
\tableline\tableline
 & Peak & Centroid & FWHM \\ 
 & (mJy beam$^{-1}$) & (km s$^{-1}$) & (km s$^{-1}$) \\
\tableline
Central & 1107.9 $\pm$ 1.6 & 453.7 $\pm$ 5.6 & 97.7 $\pm$ 0.6 \\
Blueshifted & 466.4 $\pm$ 5.6 & 379.1 $\pm$ 0.2 & 56.7 $\pm$ 0.5 \\
Redshifted & 365.9 $\pm$ 4.4 & 512.3 $\pm$ 0.1 & 28.3 $\pm$ 0.4 \\
\tableline
\end{tabular}
\tablecomments{These fits were performed for the spectra at the AGN position, 
measured with the single synthesized beam ($0\arcsec.71 \times 0\arcsec.66$). 
Results of the single Gaussian fit are summarized in Table \ref{tbl5}.} 
\end{center}
\end{table}

\section{Comparison with the Model}\label{sec6}
\subsection{Line intensities in our modeled torus}\label{sec6.1}
To interpret the physical/dynamical nature of 
the circumnuclear obscuring structures from the observed line cubes, 
we used a snapshot of our multi-phase dynamic torus model 
tuned for Circinus \citep{2016ApJ...828L..19W}, 
where non-equilibrium XDR chemistry \citep{2005A&A...436..397M} was solved, 
to predict the emission line properties of multiple species. 

The simulated CND has a circular gas distribution for simplicity with $r \simeq 16$ pc, 
with the total $M_{\rm H_2}$ of $2 \times 10^6~M_\odot$ (128$^3$ grids; 0.25 pc resolution), 
which is roughly consistent with the observed $M_{\rm H_2}$ in Circinus 
measured with a slightly larger aperture 
($\sim 3-4 \times 10^6~M_\odot$ in the $r \sim 35$ pc disk with $\alpha_{\rm CO} = 0.8$ $M_\odot$ (K km s$^{-1}$ pc$^2$)$^{-1}$; \S~4). 
The central AGN has the same $M_{\rm BH}$ and the Eddington ratio as those observed (\S~1). 
The ultraviolet radiation from the central thin accretion disk is angle-dependent, 
whereas we assumed spherically symmetric X-ray radiation from the AGN corona. 
Self-gravity of the gas is ignored as it is not essential 
for the dynamics in the fountain scheme \citep[see also][]{2012ApJ...758...66W}. 
After the XDR chemistry (abundance) was solved, 
post-processed non-LTE radiative transfer calculations were performed. 
Once the radiation field and optical depth are determined in each grid, 
we can observe line emission from an arbitrary direction \citep[for further details see][]{2018ApJ...852...88W}. 
For example, \citet{2018ApJ...852...88W} presented the resultant properties 
of multiple $^{12}$CO transition lines from $J = 0$ to 15. 
Their procedure was fully applied to the case of the \cone in this work.

\subsection{Results of our simulation}\label{sec6.2}
\begin{figure}
\begin{center}
\includegraphics[scale=0.45]{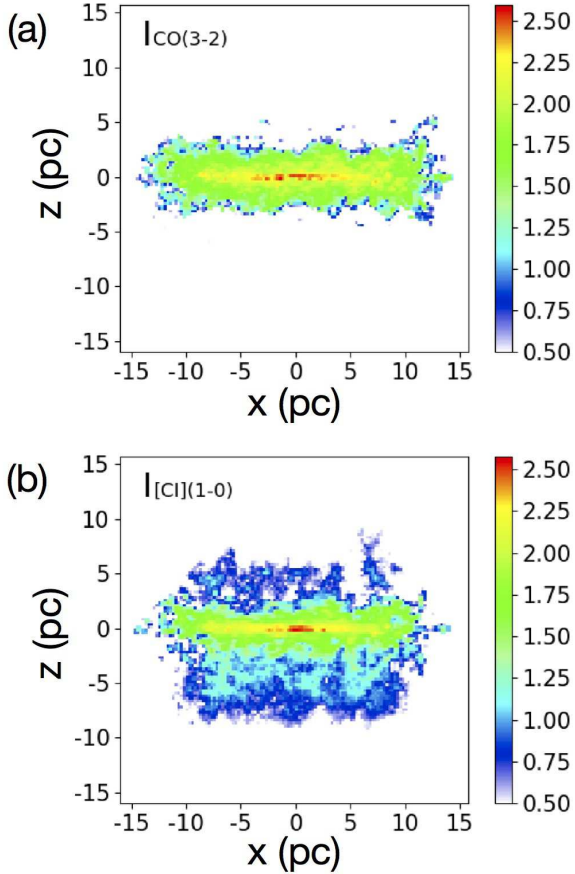}
\caption{
Simulated velocity-integrated brightness temperature maps of (a) CO(3--2) 
and (b) \cone in the central region of the Circinus galaxy 
for $i = 90\arcdeg$ (edge-on), shown on a logarithmic scale. 
A simple circular disk geometry is assumed. 
The \cone emission is spatially more extended along the $z$-direction than the CO(3--2). 
}
\label{fig19}
\end{center}
\end{figure}

Figure \ref{fig19} shows the edge-on distributions of the simulated 
integrated intensities of CO(3--2) and \cone emission lines. 
This reinforces our argument that the diffuse atomic gas traced by the \cone is more extended 
along the $z$-direction of the disk than the dense molecular gas 
traced by the CO(3--2) at this spatial scale ($r \lesssim 10$ pc), 
which is due to the atomic outflows in our model. 
On the other hand, one can recognize that the most intense part of the \cone distribution 
essentially traces the same region as the CO(3--2) distribution. 
Note that the hydrogen (H$_2$ + \ion{H}{1}) volume density in the disk mid-plane 
($\sim 10^5$ cm$^{-3}$) is $\gtrsim 100$ times higher 
than that of the spatially/vertically-extended region \citep[$\lesssim 10^3$ cm$^{-3}$,][]{2016ApJ...828L..19W}. 
Thus, the line-emitting region can be divided into the following two parts: 
\begin{itemize}
\item $|\Delta z| \lesssim 3$ pc: atomic gas and molecular gas coexist. 
Turbulence due to the failed winds determines the geometrical thickness of this region. 
The observed CO(3--2) line profile, as well as the central Gaussian component 
of the \cone line profile (Figure \ref{fig18}) reflect this component. 
\item $|\Delta z| \gtrsim 3$ pc: preferentially seen in the \cone emission, not in the CO(3--2) emission. 
The AGN-driven outflows, which are selectively seen in the ionized and atomic gas phase, 
are responsible for producing this geometrically thick structure in the \cone distribution. 
\end{itemize}

\begin{figure*}
\begin{center}
\includegraphics[scale=0.5]{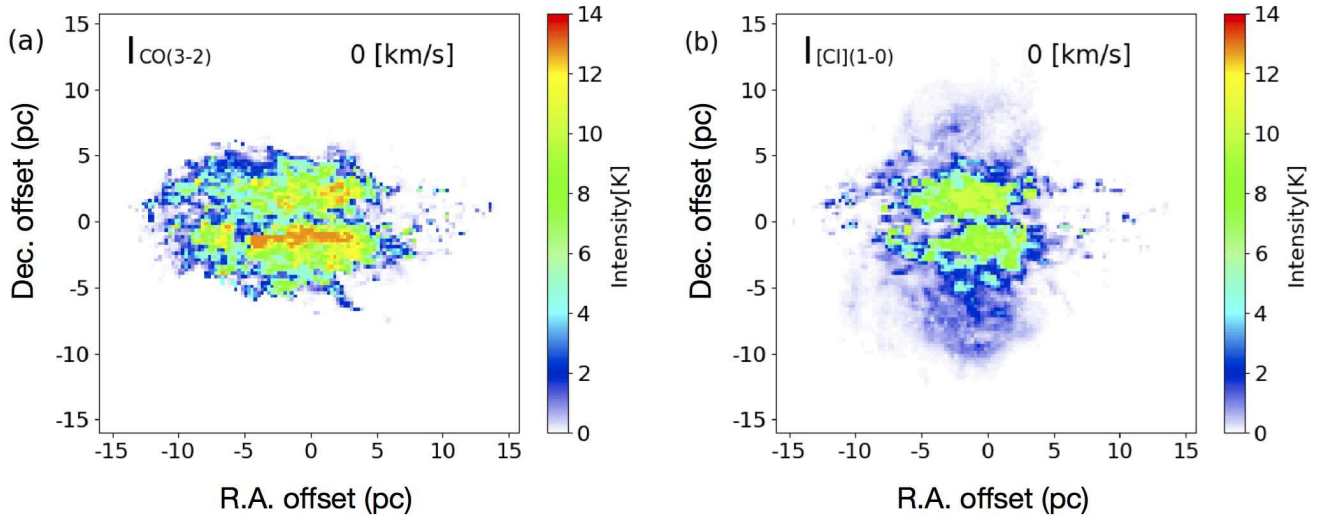}
\caption{
Simulated channel map of (a) CO(3--2) and (b) \cone intensities (in units of the brightness temperature) 
of the CND of the Circinus galaxy, with $i = 75\arcdeg$. 
Examples at the systemic velocity are shown. 
In both (a) and (b), the bulk of the emission stems from the region delineated by the CO(3--2), 
which corresponds to the geometrically thin mid-plane of the disk if viewed at $i = 90\arcdeg$ (i.e., an inclined thin disk; Figure \ref{fig19}). 
On the other hand, the spatially extended (or elongated along the vertical direction) component 
is only visible in the case of the [\ion{C}{1}](1--0), as expected from Figure \ref{fig19}. 
This geometrically thick structure is due to the AGN-driven outflows. 
}
\label{fig20}
\end{center}
\end{figure*}

Figure \ref{fig20} shows simulated channel maps of (a) CO(3--2) intensity and (b) \cone intensity 
at the systemic velocity of this galaxy with $i = 75\arcdeg$ 
\citep[the best-fit value to reproduce the IR SED of Circinus with our model;][]{2016ApJ...828L..19W}. 
For both the CO(3--2) and the [\ion{C}{1}](1--0), the bulk of the emission comes from the region outlined by the CO(3--2) distribution. 
This region corresponds to the geometrically thin mid-plane of the CND, 
if viewed from the edge-on angle (Figure \ref{fig19}); 
therefore, we call this the inclined thin disk. 
Owing to the high inclination adopted here, 
we are still able to see the spatially extended 
(or vertically elongated along the $z$-direction due to outflows) component 
besides the inclined thin disk in the \cone map (Figure \ref{fig20}b). 
As the outflows would have coherent velocities, 
we expect that blueshifted and redshifted components with such velocities can be superposed 
on a simple Gaussian profile in the simulated \cone line profile, which is confirmed in the analysis below. 

\begin{figure}
\begin{center}
\includegraphics[scale=0.38]{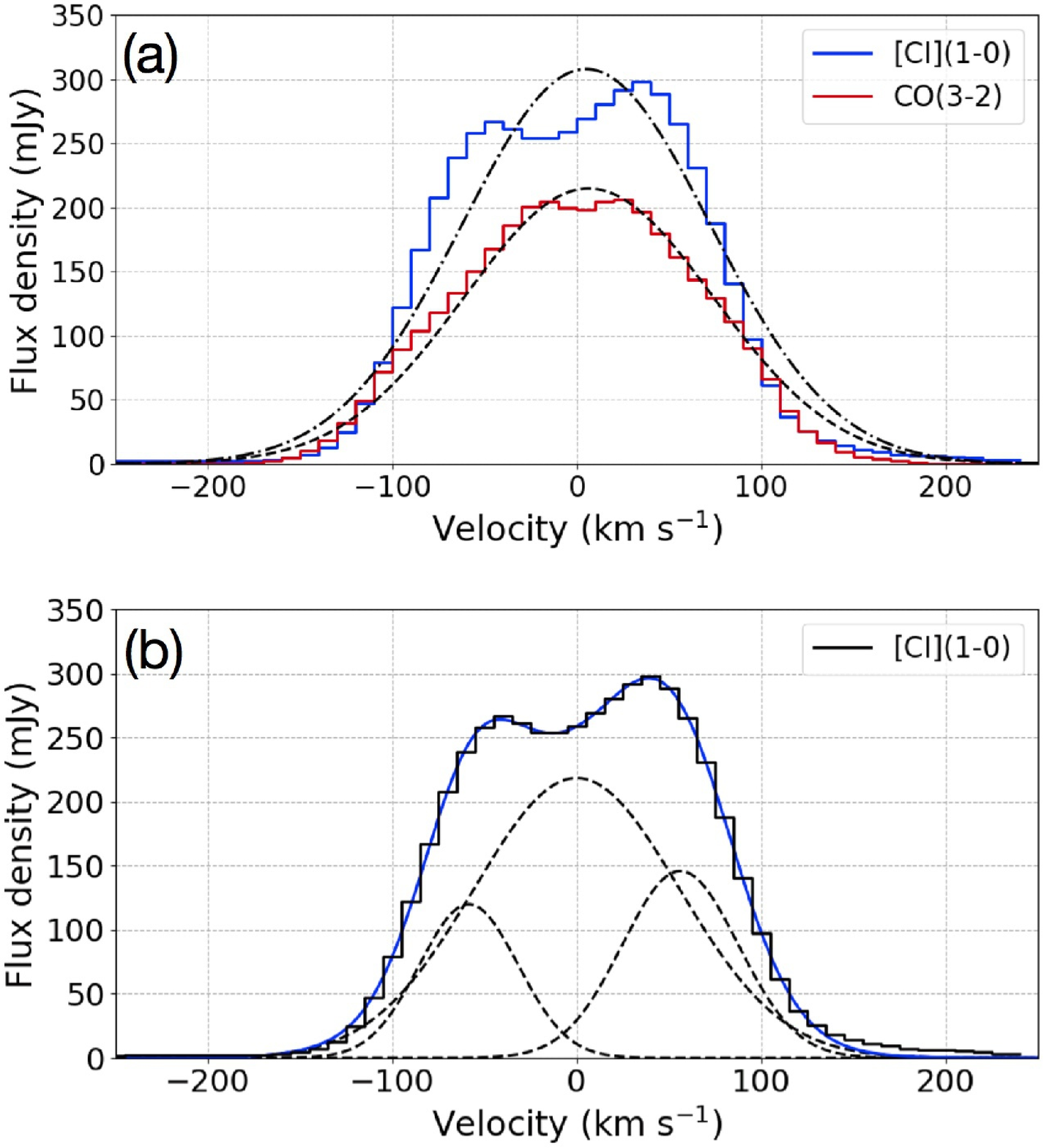}
\caption{
(a) Simulated line profiles of the \cone and the CO(3--2) emission 
measured over the entire CND (see also Figure \ref{fig19}), seen with $i = 75\arcdeg$. 
The CO(3--2) follows a single Gaussian profile well (dashed line), 
while deviation from a single Gaussian profile (dot-dashed line) is recognized in the case of the \cone. 
(b) The same \cone spectrum as in (a), 
with a triple Gaussian fit line (each component is indicated by the dashed lines). 
The blueshifted and redshifted components represent the coherent outflows. 
Results of these fits are summarized in Table \ref{tbl7}. 
}
\label{fig21}
\end{center}
\end{figure}

Figure \ref{fig21}a therefore exhibits how the \cone and CO(3--2) line profiles appear 
when viewed with $i = 75\arcdeg$ based on our model (see also Table \ref{tbl7} for the results of the Gaussian fits). 
The simulated CO(3--2) profile that basically reflects the rotating mid-plane component of the inclined thin disk 
can be explained by a single Gaussian, as expected, 
whereas deviation from a single Gaussian is evident in the case of the [\ion{C}{1}](1--0) profile. 
The \cone outflows are then clearly manifested 
in the three components Gaussian fit to the spectrum (Figure \ref{fig21}b). 
The blueshifted and redshifted features emerge at around the coherent 
outflow velocity of $\sim 50-60$ km s$^{-1}$ in our model. 
Hence, we here suggest that (i) this qualitative difference between the CO(3--2) and the \cone line profiles, 
i.e., the existence of the additional two components to the single Gaussian profile in the case of the [\ion{C}{1}](1--0), 
as well as (ii) the centroid velocities of these outflow components, 
are consistent with the observed profiles (Figure \ref{fig11}b and \ref{fig18}). 
Note that the amplitudes of the model components are smaller than, 
and the relative strengths among the three \cone components 
are somewhat different from, the observed ones (see Tables \ref{tbl6} and \ref{tbl7}). 
These discrepancies can be reconciled by adjusting some input parameters of the model 
(e.g., AGN luminosity, $M_{\rm H2}$), which are generally associated with considerable uncertainties. 
Lastly, we found that the outflow velocities of the \cone implied by our observations 
($\sim 60$ and $\sim 75$ km s$^{-1}$; Figure \ref{fig18}) 
will not significantly exceed the escape velocity ($v_{\rm esc}$) from the vicinity of this AGN
\footnote{Even if we only consider the gravitational potential induced by the central SMBH, 
we will obtain high escape velocities of $v_{\rm esc} = 120$ km s$^{-1}$ at $r = 1$ pc and 70 km s$^{-1}$ at $r = 3$ pc, for example. 
Moreover, if we indeed use the $V_{\rm rot}$ found 
at the innermost few pc region in this work ($\sim 70$ km s$^{-1}$; Figure \ref{fig16}), 
we will also obtain a rather high $v_{\rm esc}$ of $\sim 100$ km s$^{-1}$.}. 
Thus a significant fraction of the gas carried by those winds will eventually fall back to the disk as failed winds. 
{\it Therefore, given these consistencies, we conclude that 
our results support the view that a radiation-driven fountain including outflows 
indeed functions in the central region of Circinus to form a geometrically thick structure, 
which would explain the physical origin of its torus}. 

\begin{table}
\begin{center}
\caption{Results of the Gaussian fit to the model spectra \label{tbl7}}
\begin{tabular}{cccc}
\tableline\tableline
 & Peak & Centroid & FWHM \\ 
 & (mJy beam$^{-1}$) & (km s$^{-1}$) & (km s$^{-1}$) \\
\tableline
\multicolumn{4}{c}{Single Gaussian}\\ \hline
CO(3--2) & 214.7 $\pm$ 4.1 & 6.5 $\pm$ 1.0 & 158.6 $\pm$ 2.3 \\
\cone & 307.8 $\pm$ 8.7 & 5.1 $\pm$ 2.2 & 158.3 $\pm$ 5.2 \\ \hline
\multicolumn{4}{c}{Triple Gaussians to the \cone line}\\ \hline
Central & 218.4 $\pm$ 0.4 & 0$^\dag$ & 130.7 $\pm$ 0.1 \\
Blueshifted & 120.0 $\pm$ 0.3 & -58.5 $\pm$ 0.1 & 62.9 $\pm$ 0.1 \\
Redshifted & 145.9 $\pm$ 0.3 & 56.2 $\pm$ 0.1 & 74.0 $\pm$ 0.1 \\
\tableline
\end{tabular}
\tablecomments{The centroid velocities are measured with respect to the systemic velocity. 
$^\dag$The centroid of this component was fixed to be 0 km s$^{-1}$ for simplicity.} 
\end{center}
\end{table}

\section{Summary}\label{sec7} 
Our high-resolution ALMA observations of the CO(3--2) (5.9 pc $\times$ 4.9 pc resolution) 
and the \cone (14.5 pc $\times$ 13.4 pc resolution) emission lines 
toward the CND of the Circinus galaxy revealed a wealth of detail about the atomic/molecular torus. 
In particular, we compared the observed results with predictions 
based on our multi-phase dynamic torus model, 
where the circulation of outflows, failed winds, and inflows of multi-phase gas 
jointly and virtually constitute a geometrically thick structure that can replace the classic torus. 
The main findings of this work are summarized as follows: 

\begin{itemize}
\item[1.] Both Band 7 (351 GHz) and 8 (492 GHz) continuum emission was 
detected at the CND and spiral arms in the central 200 pc region of Circinus; 
these trace thermal dust emission. 
The CND appears as the inner extension of the larger scale spiral arms 
and traces the radially farther region than the warm dust distribution seen in the $Ks$/F814 flux ratio map. 
This manifests the existence of a temperature-dependent circumnuclear dust structures. 
\item[2.] The double Gaussian fit to the Band 7 continuum visibility data 
suggested a compact polar elongation at this submm wavelength, 
in addition to a bright part of the extended CND. 
The size and orientation of the submm polar elongation 
are consistent with those seen at the MIR wavelength, 
which challenges the classic torus paradigm. 
\item[3.] The CND traced by both the CO(3--2) line (74$\pm$3 pc $\times$ 34$\pm$1 pc, PA = 212$\arcdeg$) 
and the \cone line emission carries a huge amount of H$_2$ gas of $\simeq 3 \times 10^6~M_\odot$, 
with a beam-averaged column density of $N_{\rm H2} \simeq 5 \times 10^{23}$ cm$^{-2}$. 
This would contribute significantly to the Compton thickness of this AGN 
once beam dilution is corrected for, i.e., the CND plays an important role as a part of a nuclear obscurer or a torus. 
\item[4.] We found an increasing trend of the [\ion{C}{1}](1--0)/CO(3--2) line ratio 
in positions closer to the AGN. 
The ratio at the AGN position (brightness temperature scale; 14.5 pc $\times$ 13.4 pc aperture) is $\sim 0.9$, 
which is significantly higher than those observed in nearby starburst galaxies, for example. 
One possible explanation would be the XDR chemistry, 
where the AGN radiation efficiently dissociates CO molecules. 
\item[5.] We decomposed the velocity fields traced by the CO(3--2) and the \cone emission lines 
into intrinsic rotational velocities ($V_{\rm rot}$) and dispersions ($\sigma_{\rm disp}$) with tilted rings. 
The $\sigma_{\rm disp}/V_{\rm rot}$ profile (an indicator of the disk aspect ratio) 
based on the full-resolution CO(3--2) cube suggests 
that the dense molecular gas disk is geometrically thin (ratio $\sim 0.25$) at $r \gtrsim 15$ pc, 
whereas the ratio increases to $\sim 0.4$ at $r \lesssim 10$ pc. 
Within the qualitative framework of our model, the latter 
moderate thickness of the molecular disk at this spatial scale 
is consistent with a view that the material is puffed up due to the turbulence 
induced by the AGN-driven failed winds. 
\item[6.] We found a possible indication of nuclear Keplerian rotational motion 
in the full-resolution CO(3--2) data. 
As the associated uncertainties are large, however, 
higher-resolution observations are mandatory to make a firmer conclusion. 
\item[7.] The $\sigma_{\rm disp}/V_{\rm rot}$ ratio of the \cone velocity field at $r \lesssim 10$ pc is 
significantly higher than that probed by the CO(3--2) line 
(the $uv$-range and the beam sizes are matched). 
This indicates that the diffuse atomic gas is extended in a geometrically thicker structure 
than the dense molecular gas, i.e., the torus has a multi-phase dynamic structure. 
According to our model, 
outflows preferentially seen at ionized and atomic gas 
are the prime driver of this geometrically thick atomic disk at the nuclear scale. 
\item[8.] We simulated detailed CO(3--2) and \cone distributions and their line properties 
based on our multi-phase dynamic torus model. 
We showed that the \cone emission is indeed spatially 
more extended along the vertical direction of the disk than the CO(3--2) emission. 
The simulated \cone line shows a deviation from a single Gaussian profile due to nuclear outflows, 
which is indeed consistent with the observed \cone spectrum. 
The modest outflow velocity indicates that a significant fraction 
of the wind mass will fall back to the disk, i.e., failed winds, as expected in our model to function the fountain process. 
\end{itemize}

Given various consistencies between the observations and our model predictions, 
we support the validity of the radiation-driven fountain scheme at the central region of Circinus, 
which would explain the long-lasting mystery, the physical origin of the AGN torus.

\acknowledgments 
We acknowledge the anonymous referee for his/her thorough 
reading and very constructive suggestions which improved this paper significantly. 
We thank Dr. M.~Mezcua for kindly providing us the NIR $Ks$-band and {\it HST}/F814 data. 
This paper makes use of the following ALMA data: 
ADS/JAO.ALMA\#2016.1.01613.S. 
ALMA is a partnership of ESO (representing its member states), 
NSF (USA) and NINS (Japan), together with NRC (Canada), 
NSC and ASIAA (Taiwan), and KASI (Republic of Korea), 
in cooperation with the Republic of Chile. 
The Joint ALMA Observatory is operated by ESO, AUI/NRAO and NAOJ. 
This work is supported by JSPS KAKENHI Grant Number 17K14247 (T.I.), 16H03959 (K.W.), 
and the JSPS Grant-in-Aid for Scientific Research (S) JP17H06130 (K.K.).

\appendix
Velocity channel maps of the CO(3--2) and \cone emission lines are shown here. 
These exhibit the circumnuclear spatial structures, 
including the CND and the three spiral arms (see also Figures \ref{fig6} and \ref{fig13}). 

\begin{figure*}[h]
\begin{center}
\includegraphics[scale=0.4]{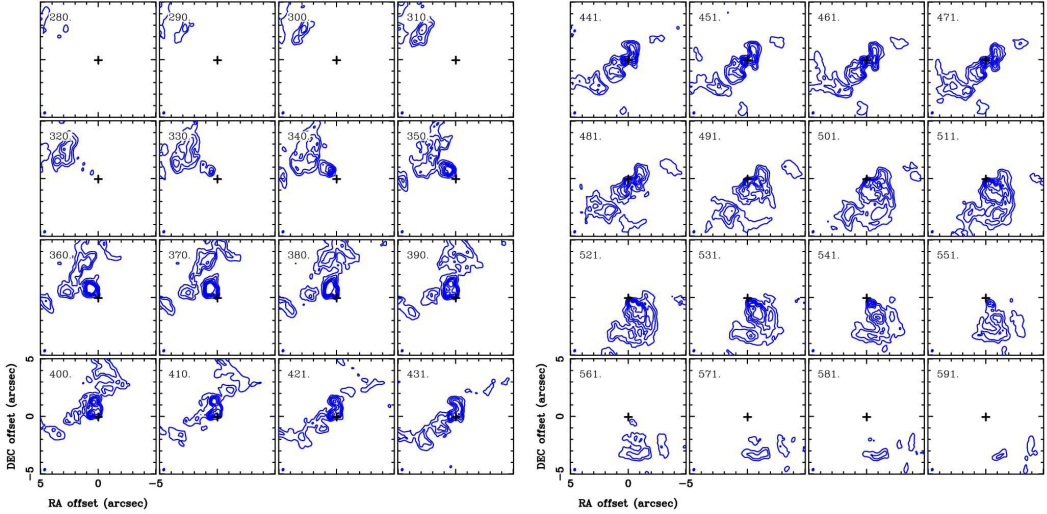}
\caption{
Velocity channel maps of the CO(3--2) emission line in the central 
$10\arcsec \times 10\arcsec$ (or $\sim 200 \times 200$ pc$^2$) box of the Circinus galaxy. 
The AGN position is indicated by the black plus sign. 
The velocity step is 10 km s$^{-1}$, and the central velocity of each channel is plotted in the upper left corner. 
Only bright emission are shown here (100, 200, 300, 400, 500, and 600$\sigma$; 1$\sigma$ = 0.37 mJy beam$^{-1}$), 
and negative contours are not shown, to enhance the clarity of the structure. 
Note that the synthesized beam size in these maps is $0\arcsec.29 \times 0\arcsec.24$ with PA = $153\arcdeg.6$. 
}
\label{figA_1}
\end{center}
\end{figure*}

\begin{figure*}[h]
\begin{center}
\includegraphics[scale=0.4]{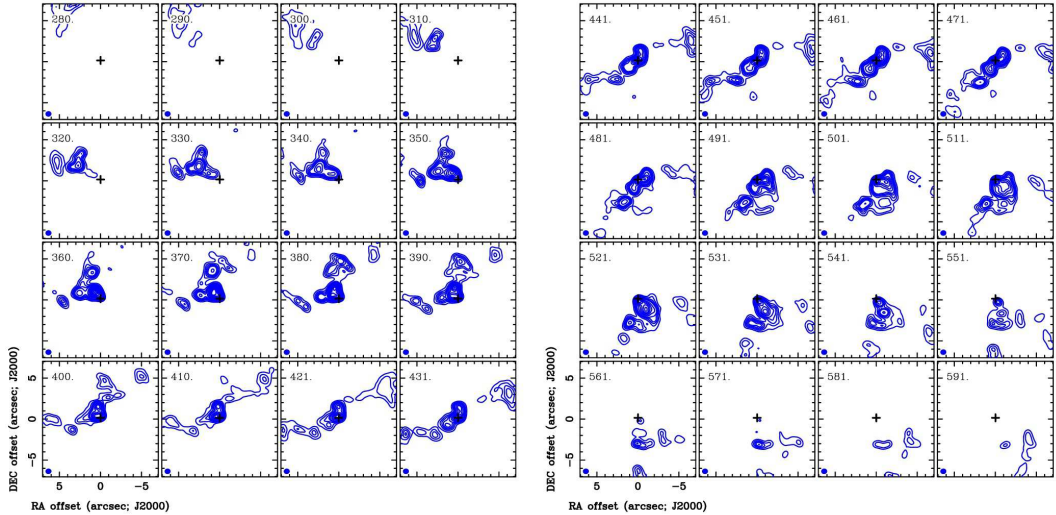}
\caption{
Velocity channel maps of the \cone emission line in the central 
$10\arcsec \times 10\arcsec$ (or $\sim 200 \times 200$ pc$^2$) box of the Circinus galaxy. 
The AGN position is indicated by the black plus sign. 
The velocity step is 10 km s$^{-1}$, and the central velocity of each channel is plotted in the upper left corner. 
Only bright emission are shown here (20, 40, 60, 80, 100, 150, 200, 250, 300, 350, 400 and 450$\sigma$; 1$\sigma$ = 3.1 mJy beam$^{-1}$), 
and negative contours are not shown, to enhance the clarity of the structure. 
Note that the synthesized beam size in these maps is $0\arcsec.71 \times 0\arcsec.66$ with PA = $95\arcdeg.9$. 
}
\label{figA_2}
\end{center}
\end{figure*}

\end{document}